\documentclass[aps,prl,twocolumn,floatfix,footinbib,showpacs,superscriptaddress]{revtex4-1}
\usepackage{graphicx}
\usepackage{amsfonts,amsmath,amssymb}
\usepackage{amsthm}
\usepackage{dsfont,bm}
\usepackage{color}
\usepackage{soul} 
\usepackage{amsbsy}
\usepackage[colorlinks=true,linkcolor=blue,pagecolor=blue,filecolor=blue,menucolor=blue,urlcolor=blue,citecolor=blue,anchorcolor=blue]{hyperref}

\usepackage{mathbbol}

\begin{document}

\title{ Comparison of Optical Response from DFT Random Phase Approximation and Low-Energy Effective Model: Strained Phosphorene}

\author{Mohammad Alidoust}
\affiliation{Department of Physics, Norwegian University of Science and Technology, N-7491 Trondheim, Norway}
\author{Erlend E. Isachsen}
\affiliation{Department of Physics, Norwegian University of Science and Technology, N-7491 Trondheim, Norway}
\author{Klaus Halterman}
\affiliation{Michelson Lab, Physics Division, Naval Air Warfare Center, China Lake, California 93555, USA}
\author{Jaakko Akola} 
\affiliation{Department of Physics, Norwegian University of Science and Technology, N-7491 Trondheim, Norway}
\affiliation{Computational Physics Laboratory, Faculty of Natural Sciences, Tampere University of Technology, FI-33101 Tampere, Finland}

\begin{abstract}
The engineering of the optical response of materials is a paradigm that demands microscopic-level accuracy
 and reliable predictive theoretical tools. Here we compare and contrast the 
 dispersive  permittivity tensor, using both a low-energy effective model and density functional theory (DFT). 
As a representative material, phosphorene subject to strain is considered. 
Employing a low-energy model Hamiltonian with a Green's function current-current correlation function, 
we compute the dynamical optical conductivity and its associated permittivity tensor.
For the DFT approach, first-principles calculations make use of 
the first-order random-phase approximation. 
Our results reveal that although the two models are generally in agreement within the low-strain and 
low-frequency regime, 
the intricate 
 features associated with the fundamental physical properties of the system and optoelectronics device implementation such as band gap, Drude absorption response,
  vanishing real part, absorptivity, and sign of permittivity over the frequency range show significant discrepancies.
  Our results suggest that the random-phase approximation employed in widely used DFT packages should be 
   revisited and improved to be able to predict these fundamental 
   electronic characteristics of a given material with confidence. Furthermore, employing the permittivity results from both models, we uncover the pivotal role that phosphorene can play in optoelectronics devices to facilitate highly programable  perfect absorption of electromagnetic waves by manipulating the chemical potential and exerting strain and illustrate how reliable predictions for the dielectric response of a given material are crucial to precise device design.             
\end{abstract}

\date{\today}

\maketitle

\section{introduction}

The dynamical finite-frequency optical conductivity and permittivity are the most 
pivotal quantities in designing optoelectronics devices.\cite{R.Gutzler,F.N.Xia} 
Various measurable optical properties such as the complex index of refraction, 
the reflectivity, and absorptivity are governed directly by the permittivity of the medium, 
which in turn is directly related to the optical conductivity of a time varying incident electromagnetic 
(EM) wave.\cite{R.Gutzler,T.Ahmed} 
The permittivity also connects the mutual influence of the medium and  electric field  of the incident EM wave, 
i.e., the light-matter interactions, and can reveal the precise nature of the medium.\cite{R.Gutzler,A.Rodin} 
Moreover, there is an emerging need for optoelectronic chip architectures that require precise
 determination and 
  manipulation of the permittivity and optical conductivity to benefit current fabrication techniques and 
  advance technological applications.\cite{F.N.Xia} Only then can fast, ultracompact low-power applications be efficiently realized.

First-principles calculations reside at the frontier 
of accurate simulations of various materials platforms. \cite{A.Carvalho}
 For example, density functional theory (DFT) calculations have shown success
 in simulating the general band gap trend as a function of the number of 
  layers in certain  two-dimensional (2D) materials, such as black phosphorus,
  which 
  constitutes
   a designed material (with desirable key characteristics such as epsilon-near-zero response \cite{S.Biswas,Alidoust2020:PRB1}), 
   and observed in experiments. \cite{V.Tran,Y.Wei,T. Fang,Z.Zhang,S.Das,Z.Qin,A.Carvalho1,G.Zhang} 
   Many 2D materials 
   consist of 2D layers of strongly bonded atoms attached to each other in the third dimension by weak forces. 
   These weakly interacting 2D layers allow for designing novel materials with
    controllable electronics characteristics 
    with low-cost operations, such as layer displacement. 
    Nevertheless, it has proven that differing functionals and approximations in DFT calculations 
    can modify the absolute band-gap of materials. 
    This issue is more pronounced in 2D materials where 
    both strong covalent bonds and weak van der Waals (vdW) forces are present.
    This is an important point in the context of DFT, which is
   unable to properly account for vdW forces without incorporating specific corrections. \cite{A.Carvalho,M.Dion,S.Grimme}

\begin{figure}[t!]
\includegraphics[width=0.40\textwidth]{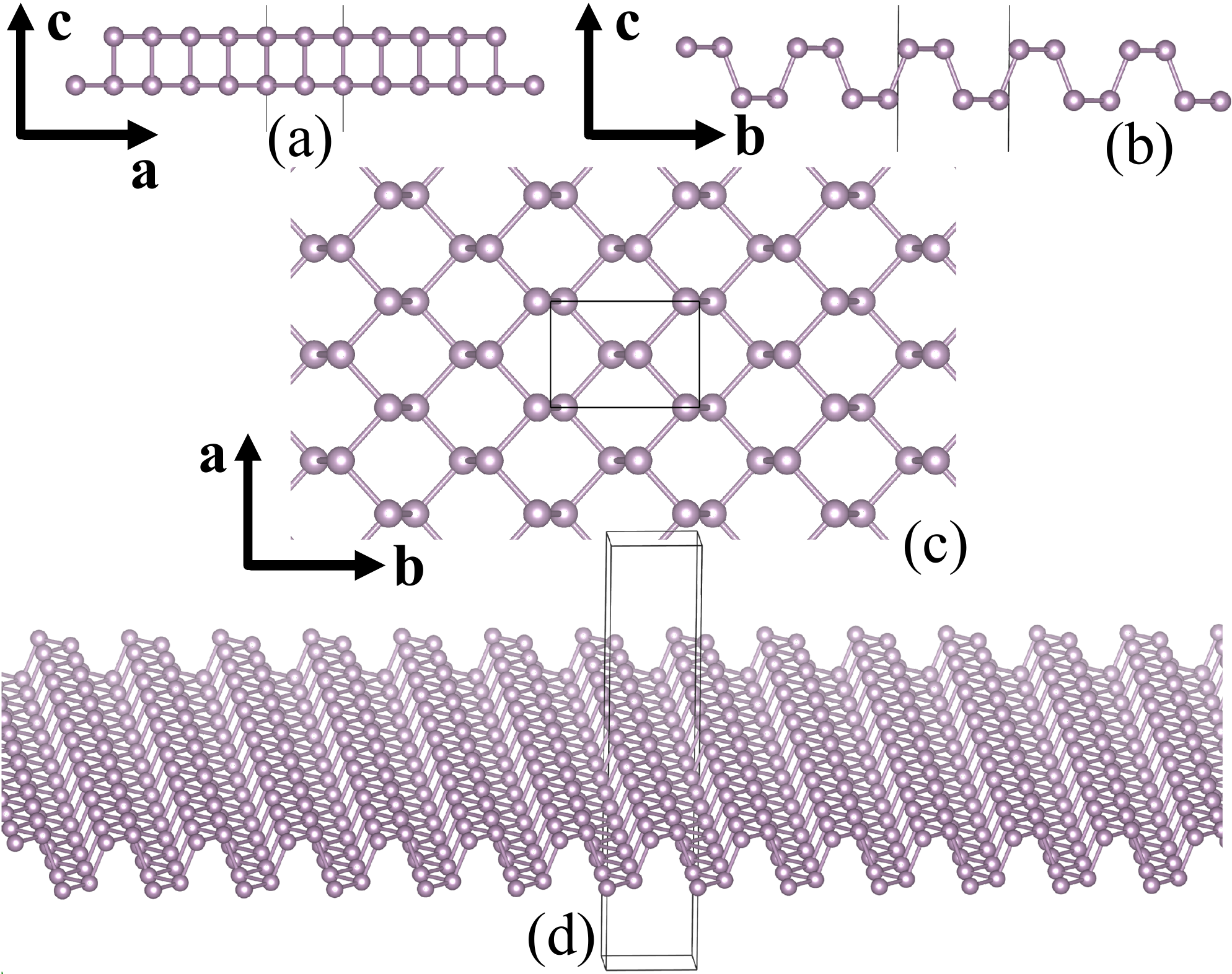}
\caption{\label{fig1} An expanded unit cell of phosphorene. (a) and (b) exhibit the side views of the 2D material along the \textbf{a} and \textbf{b} principal axes, 
whereas (c) shows phosphorene from the top view along the \textbf{c}-axis. 
Panel (d) displays the expanded crystal with a generic view. 
The 2D material is located at the middle of the unit cell with sufficiently 
large vacuum regions along the \textbf{c} principal axis. }
\end{figure}

The weak interlayer vdW interactions provide a unique opportunity 
to peel off the layers and eventually create a
one-atom-thick 2D sheet
 with drastically different electronics properties than the bulk material. 
 Furthermore, performing mechanical operations
 such as the exertion of strain,
  on a single-layer material is much easier,
 as it responds more effectively to these operations compared to
 the bulk material. 
 The most famous examples include graphene (a single layer of carbon atoms extracted from graphite)\cite{A.H.CastroNeto1} and phosphorene (shown in Fig. \ref{fig1}, a single layer of phosphorus atoms extracted from black phosphorus)\cite{H.Liu}. Unlike graphene where carbon atoms reside in a single plane, phosphorene atoms reside within two planes with a finite separation distance, making a puckered structure [see Figs.~\ref{fig1}(a) and \ref{fig1}(b)]. 
Compared to bulk black phosphorus, phosphorene acquires a fairly large band gap, $\sim 1.52$~eV, very suitable for semiconductor and field-effect transistor technologies. \cite{A.Carvalho1}
In the following, we specifically focus on phosphorene (with the possibility of incorporating strain) as its low energy Hamiltonian is available and provides an excellent semiconductor platform 
 for strictly comparing and contrasting the results of DFT and those obtained by the low energy model.   
 
As the influence of vdW forces in a single layer of black phosphorus 
weakens, 
one may expect that the deficiencies
 in the various DFT simulations described earlier would  consequently diminish. 
 However, as we shall see below, DFT with a widely-used functional still underestimates the band gap of phosphorene. 
 On the other hand, a low-energy effective model can incorporate a proper band gap,
  as it is calibrated 
  through band structure calculations
  and experimental inputs
  when parametrizing a particular model. 
  Furthermore, the low-energy effective model can provide 
  precise and deep insights
   into the fundamental physical properties of the material, such as 
  dominant transitions across the band gap, which are inaccessible in purely
   computational approaches like DFT.                   

In this paper, we compute
each component of the permittivity
tensor of phosphorene subject to in-plane strain.
Two methods are used:                                 
 One involves
  DFT combined with a random phase approximation (DFT-RPA),\cite{RPA,Sauer,E.vanLoon,M.Gajdos,E.Sasioglu,H.Shinaoka,C.Honerkamp,X.J.Han}
   and the other uses a low-energy model 
  Hamiltonian  with the
  Green's function current-current correlator. 
  Our results reveal that the permittivity components calculated from 
   DFT-RPA indicate 
  an
  anisotropic band gap (direction dependent) with  magnitude 
  that is incompatible  with
   the corresponding band structure obtained from 
    DFT and 
   the
   Perdew-Burke-Ernzerhof (PBE) functional. 
   The permittivity tensor computed by the low-energy model, however,
    is fully consistent with 
    the associated band gap and clearly describes the underlying 
    physical characteristics of phosphorene. 
    The low energy model  also 
    allows for studying the influence of 
    chemical potential or doping variations. 
    We show that in addition to 
    chemical potential variations, the application of strain provides an effective 
    on/off switching mechanism for the Drude response. The underlaying mechanism is the 
    on/off switching of the intraband transitions that can provide valuable information on the band structure of the system. It should be emphasized that although we have studied a specific 2D semiconductor, our conclusions are 
     generalizable to other materials and point to the urgent need for revisiting 
     DFT-RPA implementations used in many DFT packages. Finally, employing the permittivity data from the
     DFT and low-energy models, 
     we demonstrate how their differing predictions can influence the precise design of an optoelectronics device. Nevertheless, our findings with both DFT-RPA and low-energy model reveal perfect absorption of electromagnetic waves in layered devices containing phosphorene, which is highly tunable by the chemical potential of phosphorene and the application of strain to the plane of phosphorene.

The paper is organized as follows. In Sec.~\ref{formalism}, the formalisms used in both 
the DFT-RPA and low-energy models
are summarized. In Sec.~\ref{bandgap}, the components of the 
permittivity tensor will be presented and the associated physics will be analyzed through band 
structure
diagrams. 
It will be discussed how the inaccurate results of DFT-RPA 
are unable to provide correct information on the microscopic properties of the system. 
In Sec.~\ref{Drude}, the Drude absorption response,
and how it provides information on the band structure will be analyzed and discussed. 
In Sec.~\ref{device}, the results of 
DFT and low-energy models will be contrasted
in a 
practical device scenario,
 where the importance of accurate permittivity predictions are paramount
 to the  proper design of a functional optical device. 
 Finally, a summary and concluding remarks will be given in Sec.~\ref{conclusions}.  

\section{frameworks and formalisms}\label{formalism}

Below, in  Secs.~\ref{DFT} and \ref{lowEHamil}, the basics
 of the two approaches, i.e., first-principles DFT-RPA and 
 the
 low-energy effective model Hamiltonian used in the Green's function current-current correlator
  are summarized.

\subsection{First-principles density functional theory}\label{DFT}

The density functional calculations are based upon
the charge density response to sufficiently weak external interactions,
 such as an electric field. 
 In this case, the Kohn-Sham equations can be evaluated to obtain the dielectric response of the
  material. \cite{AB_dielect} 
  In most DFT packages, such as GPAW, QUANTUM ESPRESSO, and VASP, 
  a random phase approximation (RPA)\cite{RPA,Sauer,E.vanLoon} is implemented to evaluate the dielectric response,
  or permittivity tensor.\cite{M.Gajdos,gp3} Unfortunately,
this approximation neglects the exchange-correlation contribution and can 
lead to {\it unphysical} modifications to the original band structure 
obtained through a specific functional and its associated exchange-correlation. \cite{M.Gajdos} 
In the linear response regime, the dielectric matrix is given by 
\begin{equation}
\epsilon_{G,G'}(\textbf{q},\omega) = \delta_{G,G'} - \frac{4\pi}{|\textbf{q}+\textbf{G}|}\chi^0_{G,G'}(\textbf{q},\omega),
\end{equation}
which is linked to the first-order density 
response $\chi^0$, Bloch vector of the incident wave \textbf{q}, reciprocal lattice vectors \textbf{G}, and the conventional Kronecker delta $\delta_{ij}$. In the RPA regime, 
the dielectric function is obtained at the $\Gamma$ point so that 
 \begin{equation}
\epsilon(\textbf{q},\omega) = \frac{1}{\epsilon_{0,0}^{-1}(\textbf{q},\omega)}.
\end{equation}
  
In this work, first-principles DFT calculations of the dielectric response are performed using RPA 
as implemented 
in the $\rm GPAW$ DFT package. \cite{gp1,gp2,gp3} The gradient-corrected functional by PBE is used for the 
exchange-correlation 
energy when calculating the electronic band structure and the dielectric response. 
To grid ${\bf k}$ space on the basis of the 
Monkhorst-Pack scheme, a sufficiently large value, i.e., $6.0$~${\bf k}$-points per $\text{\AA}^{-1}$ is incorporated.
 The cut-off for the kinetic energy of the
 plane-waves is set to $800$~eV,
  and 60 unoccupied electronic bands are included with a convergence on the first 50 bands. 
  These high values ensure avoiding any artificial effects due to the application of strain in the subsequent calculations 
  that follow. A small imaginary part is added to the frequency variable throughout the calculations, 
  i.e., $\eta=0.01$~eV, and the width of the Fermi-Dirac distribution is fixed at $0.01$~eV.  

We introduce the strain parameters $s_{ii}$ (for the $i=x,y,z$ directions),
to describe 
 the expansion and compression of the atom's 
 location and unit cell with respect to the 
 relaxed unit cell in each  direction, i.e., 
 $a=s_{xx}  a_0$,
$b = s_{yy}  b_0$, and
$c =s_{zz} c_0$.
Here
 $a$, $b$, and $c$ are the three strained unit cell axis lengths, and the unstrained unit cell axis 
lengths are $a_0$, $b_0$, and $c_0$. 
Therefore, in this notation, $s_{xx}=s_{yy}=s_{zz}=0.9, 1.0,$ and $1.1$ correspond to 
strains of
$- 10\%$, $0\%$, and $+ 10\%$, respectively. 
The strain-free expanded unit cell with differing view angles is shown in Fig.~\ref{fig1}. 
The phosphorene sheet is located in the \textbf{a}-\textbf{b} plane and a large vacuum region
is included
 in the unit cell in the \textbf{c} direction. 
 Since
  periodic boundary conditions in all directions are set in the numerical
   simulations, the  vacuum 
   spacing in the \textbf{c} direction 
   ensures zero overlap
    of the wave-functions in replicated sheets in the \textbf{c} direction. 
 Additionally, as the system is non-magnetic, the permeability is isotropic and can be
 set to its vacuum value. 

\subsection{Low-energy effective model}\label{lowEHamil}

To study the permittivity of phosphorene 
subject to an
in-plane strain $\varepsilon_{ii}$ within the 
effective 
low-energy  regime, 
we employ the model Hamiltonian presented in Refs.~\onlinecite{Voon1,Voon2}:
\begin{eqnarray}\label{Hamil}
H= &&\int \frac{d\textbf{k}}{(2\pi)^2}\hat{\psi}^\dag_{\textbf{k}}H(\textbf{k})\hat{\psi}_{\textbf{k}}=\nonumber\\ &&\int \frac{d\textbf{k}}{(2\pi)^2}\hat{\psi}^\dag_{\textbf{k}}\Big\{ \big [u_0+\alpha_is_{ii} +(\eta_j+\beta_{ij}s_{ii})k_j^2\big]\tau_0 +\nonumber\\ &&\big[\delta_0+\mu_is_{ii} +(\gamma_j+\nu_{ij}s_{ii})k_j^2\big ]\tau_x -\chi_y k_y\tau_y\Big\}\hat{\psi}_{\textbf{k}}, 
\end{eqnarray}
where the 
indices ($i,j$) run over the
coordinates $x,y$. Here $\tau_i$ are the Pauli matrices in pseudospin space (atomic sites), 
and $\textbf{k}=(k_x,k_y)$ is the momentum. The field operator associated with the 
Hamiltonian is given by $\hat{\psi}^\dag(\textbf{k})=(\psi_{A}^\dag,\psi_{B}^\dag)$, where the pseudospins are labeled by $A$ and $B$. 
The parameters used for
 this model are summarized in Table~\ref{table}. 
 This model has also been employed to study superconductivity and supercurrent
  in strained and magnetized phosphorene systems\cite{Alidoust2018:PRB1,Alidoust2018:PRB2} 
 where  it was found that strain can induce Majorana zero energy modes and drive $s$-wave and $p$-wave superconducting correlations to $d$-wave and $f$-wave correlations that might explain experimental observations in these contexts\cite{Alidoust2019:PRB1}.

\begin {table}[t]
\caption {Band parameters of  phosphorene subject to an in-plane strain \cite{Voon1,Voon2}. } \label{table}
\begin{center}
\begin{tabular}{c*{4}{c}c}
\hline
\hline
$u_0$(eV) & $\delta_0$(eV) & $\alpha_x$(eV) & $\alpha_y$(eV) & $\mu_x$(eV)    \\
-0.42  & +0.76  & +3.15  & -0.58  & +2.65     \\
\hline
 $\mu_y$(eV) & $\eta_x$(eV$ \textup{\AA}^2$)  & $\eta_y$(eV$ \textup{\AA}^2$)  & $\gamma_x$(eV$ \textup{\AA}^2$)  & $\gamma_y$(eV$ \textup{\AA}^2$) \\
 +2.16  &  +0.58  & +1.01 & +3.93 & + 3.83  \\
\hline
$\beta_{xx}$(eV$ \textup{\AA}^2$) & $\beta_{yx}$(eV$ \textup{\AA}^2$) & $\beta_{xy}$(eV$ \textup{\AA}^2$) & $\beta_{yy}$(eV$ \textup{\AA}^2$)  \\
-3.48 & -0.57 & +0.80 & +2.39\\
\hline
$\nu_{xx}$(eV$ \textup{\AA}^2$)  & $\nu_{yx}$(eV$ \textup{\AA}^2$) & $\nu_{xy}$(eV$ \textup{\AA}^2$) & $\nu_{yy}$(eV$ \textup{\AA}^2$) & $\chi_y$(eV$ \textup{\AA}$)  \\
-10.90 & -11.33 & -41.40 & -14.80 & +5.25\\
\hline
\hline
\end{tabular}
\end{center}
\end{table}

\begin{figure*}[t!]
\includegraphics[clip, width=0.92\textwidth]{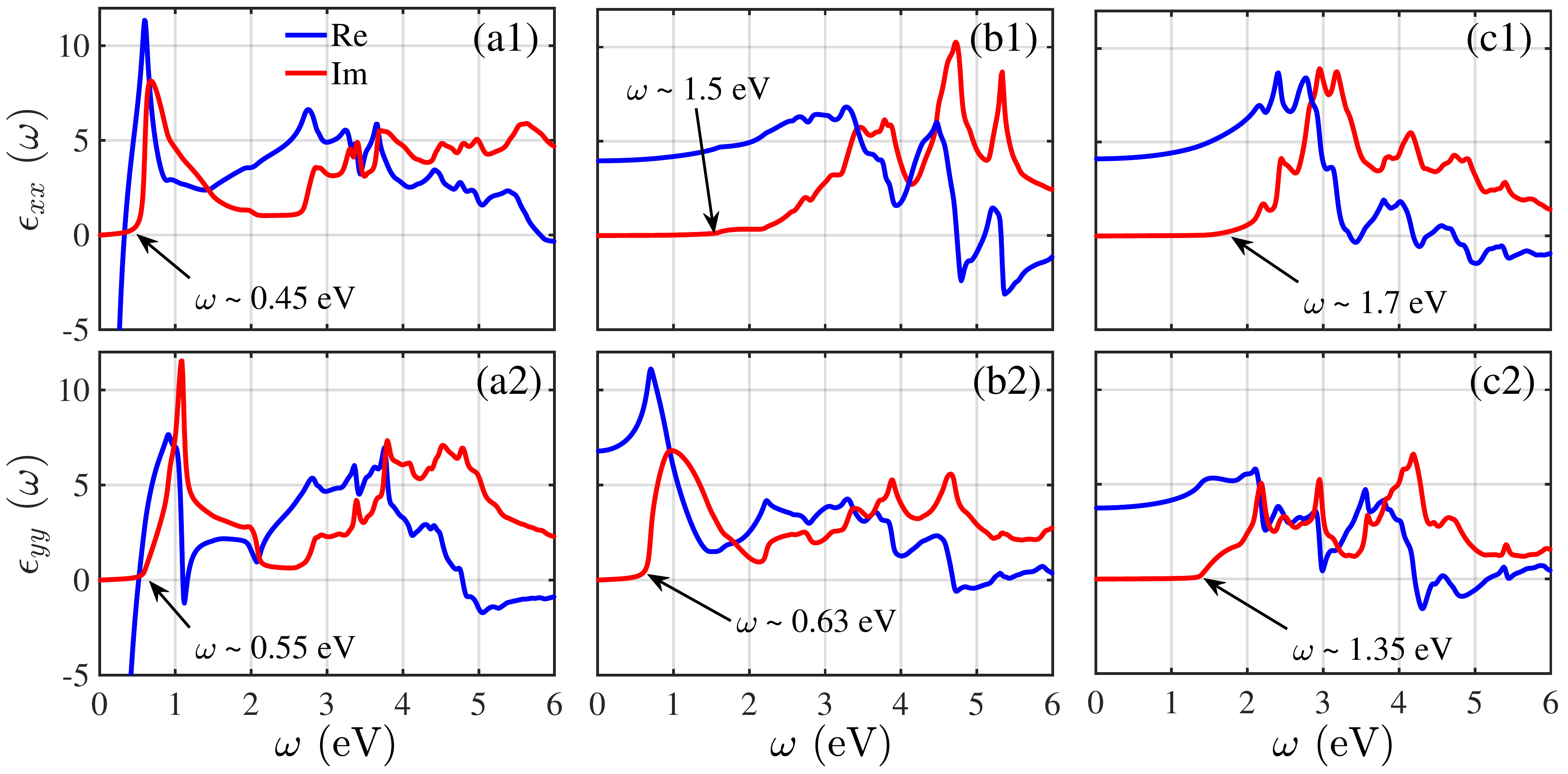}\\
\includegraphics[clip, width=0.30\textwidth]{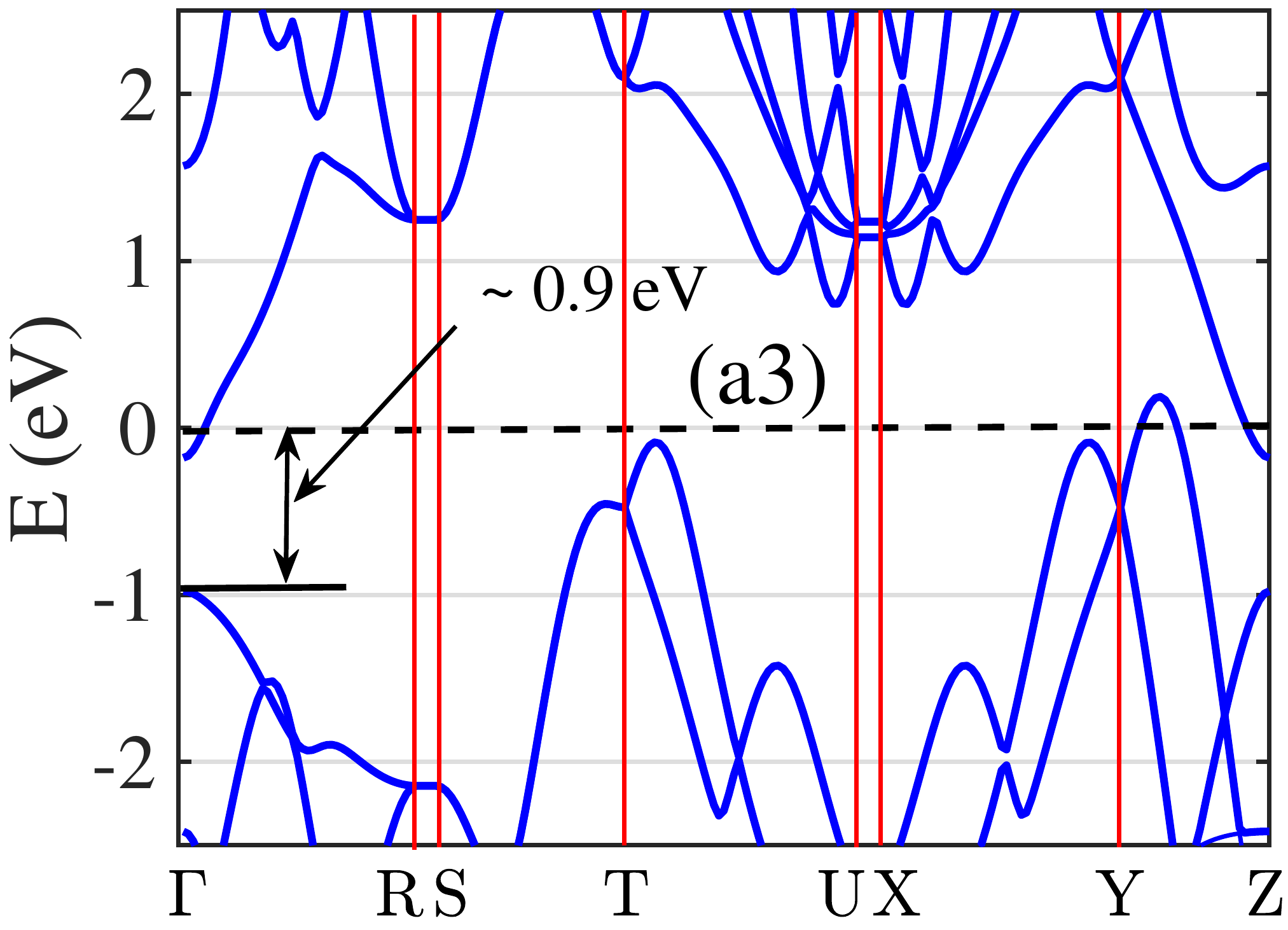}
\includegraphics[clip, width=0.30\textwidth]{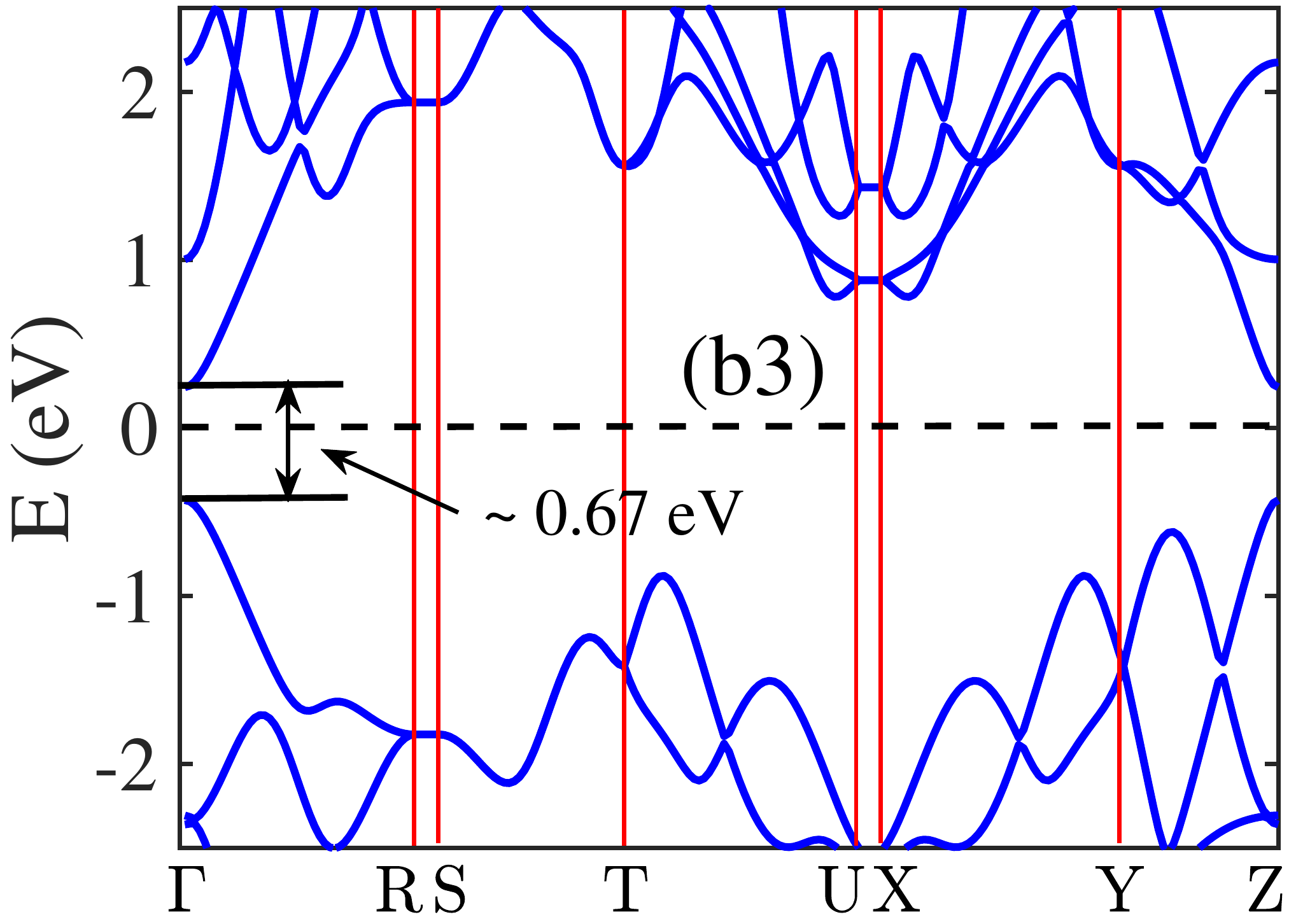}
\includegraphics[clip, width=0.30\textwidth]{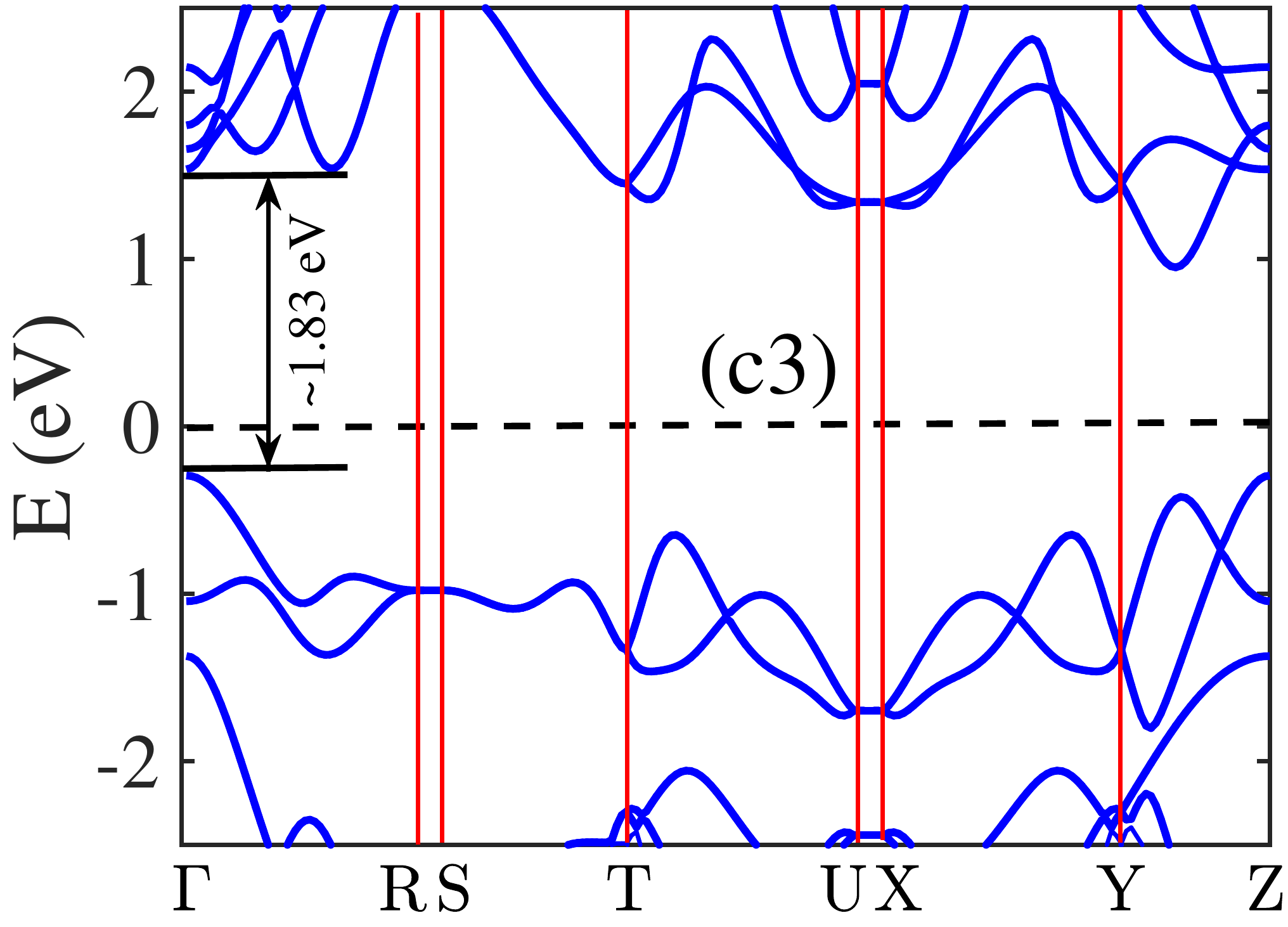}
\caption{\label{fig2} Real (blue) and imaginary (red) parts of permittivity obtained from first-principles calculations in combination with the RPA (GPAW). The top and middle rows show $\epsilon_{xx}(\omega)$ and $\epsilon_{yy}(\omega)$, respectively, whereas the bottom row is the band structure of phosphorene along the different paths in $\textbf{k}$ space. Column-wise, (a)-(c) correspond to 
biaxial strains of
$-10\%$, $0\%$, and $+10\%$, respectively. }
\end{figure*}

In the low-energy regime, the many-body dielectric response can be expressed by\cite{Mahan},
\begin{equation}\label{perm}
\epsilon_{ab}(\omega) = \delta_{ab}-\lim_{|\textbf{q}|\to 0} \frac{\Pi_{ab}(\omega,\textbf{q})-\Pi_{ab}(0,\textbf{q})}{\epsilon_0\omega^2},
\end{equation}  
in which $\delta_{ab}$ is the Kronecker-delta and $\epsilon_0$ is the vacuum permittivity. 
The current-current correlation functions are given by:
\begin{align}
\Pi_{ab}(\omega,\textbf{q}) = & e^2T\sum_n\sum_s \text{Tr} \int \frac{d^2p}{(2\pi)^2} J_{a}^s G_s(\varepsilon_n+\omega_k,\textbf{p}+\textbf{q})\nonumber\\
& \times J_b^s G_s(\varepsilon_n,\textbf{p})\Big |_{i\omega_k\to \omega + i\delta}.
\end{align} 
Here $J_{a,b}^s$ are the components of 
the current operators in
the  $a,b$ directions. 
The components of the
Green's function are labeled $G_s$, and $\omega_k=2\pi Tk$ and $\varepsilon_n=\pi T(2n+1)$ are the bosonic and fermionic Matsubara frequencies, respectively  ($k,n$ are integers). 
Finally, 
the finite-frequency optical conductivity tensor can be obtained from,
\begin{equation}\label{optcond}
\sigma_{ab}(\omega)=
\frac{i}{\omega}\lim_{|\textbf{q}|\to 0} \Big\{ \Pi_{ab}(\omega,\textbf{q})-\Pi_{ab}(0,\textbf{q})\Big\}.
\end{equation}
Below, we  employ separately these two frameworks  discussed
in  Secs.~\ref{DFT} and \ref{lowEHamil}
 and compute the components of 
the permittivity tensor.

\section{results and discussions}\label{results}
This section is divided into three subsections:
In Sec.~\ref{bandgap}, the various aspects of permittivity and their underlying
physical origins will be analyzed by visualizing the associated band structures. In Sec.~\ref{Drude}, the Drude absorption of strained phosphorene will be discussed.\cite{M.Tahir,J.Jang,C.H.Yang,D.Q.Khoa} In Sec.~\ref{device}, the device implications will be presented.

\begin{figure*}[t!]
\includegraphics[clip, width=0.92\textwidth]{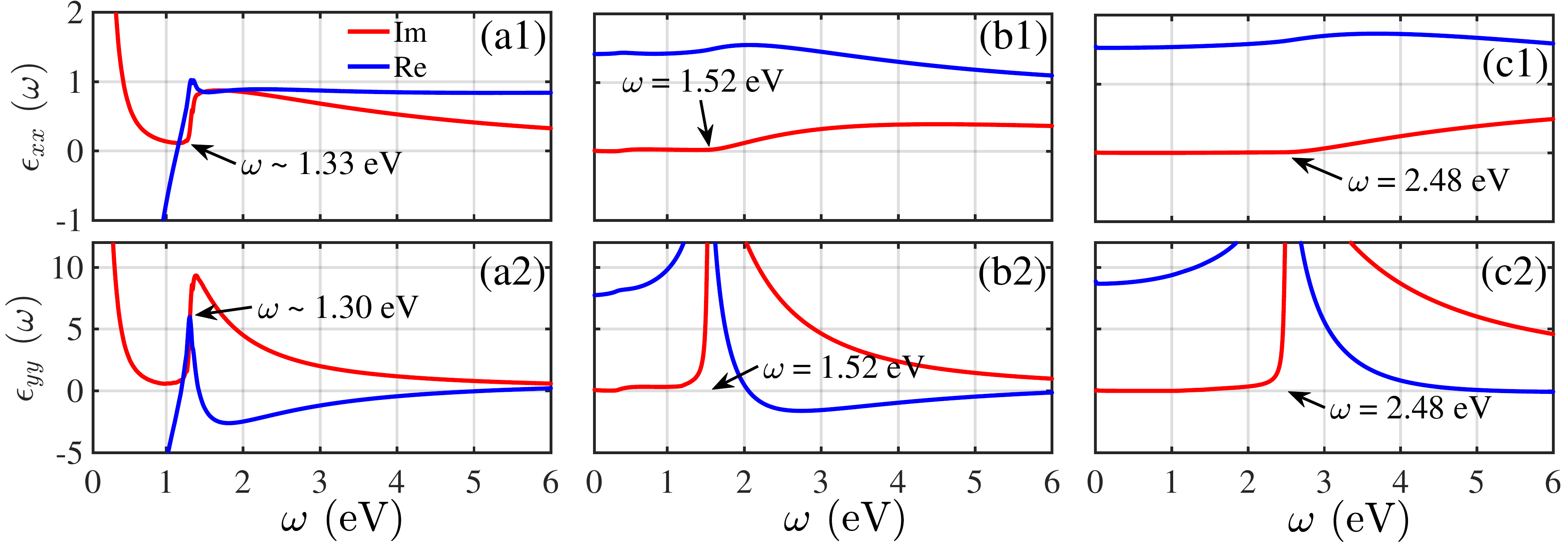}\\
\includegraphics[clip, width=0.3\textwidth]{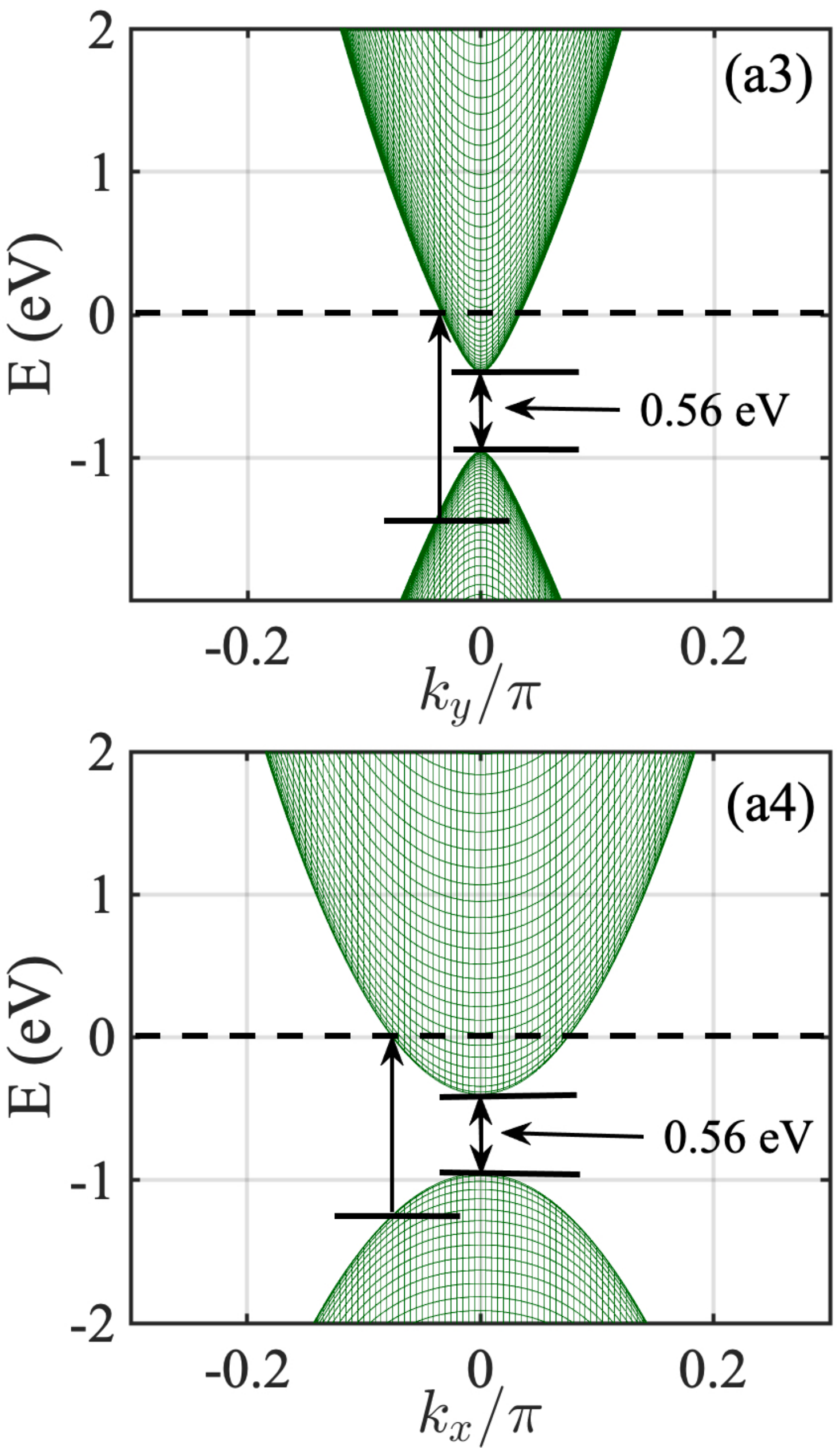}
\includegraphics[clip, width=0.3\textwidth]{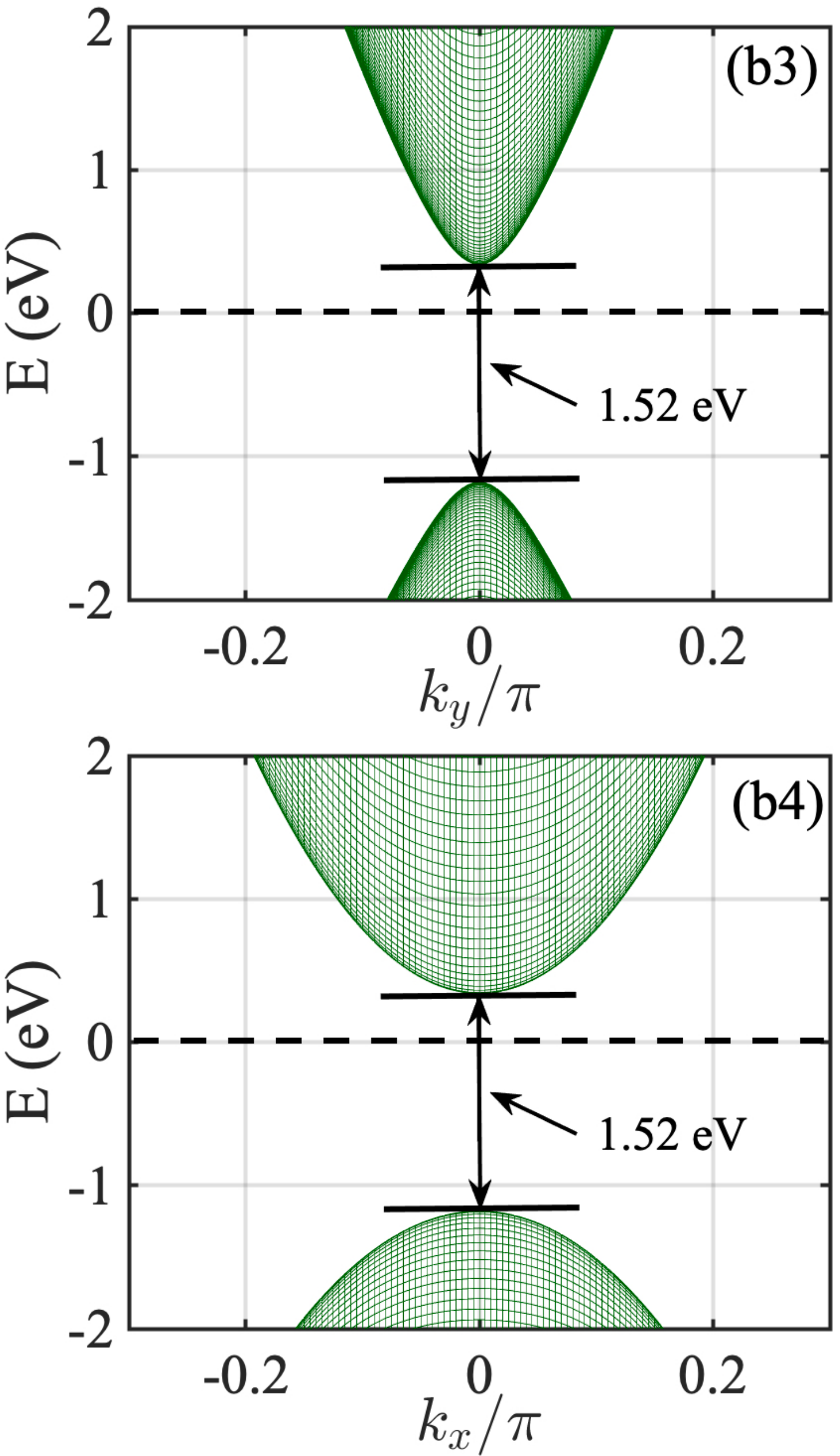}
\includegraphics[clip, width=0.3\textwidth]{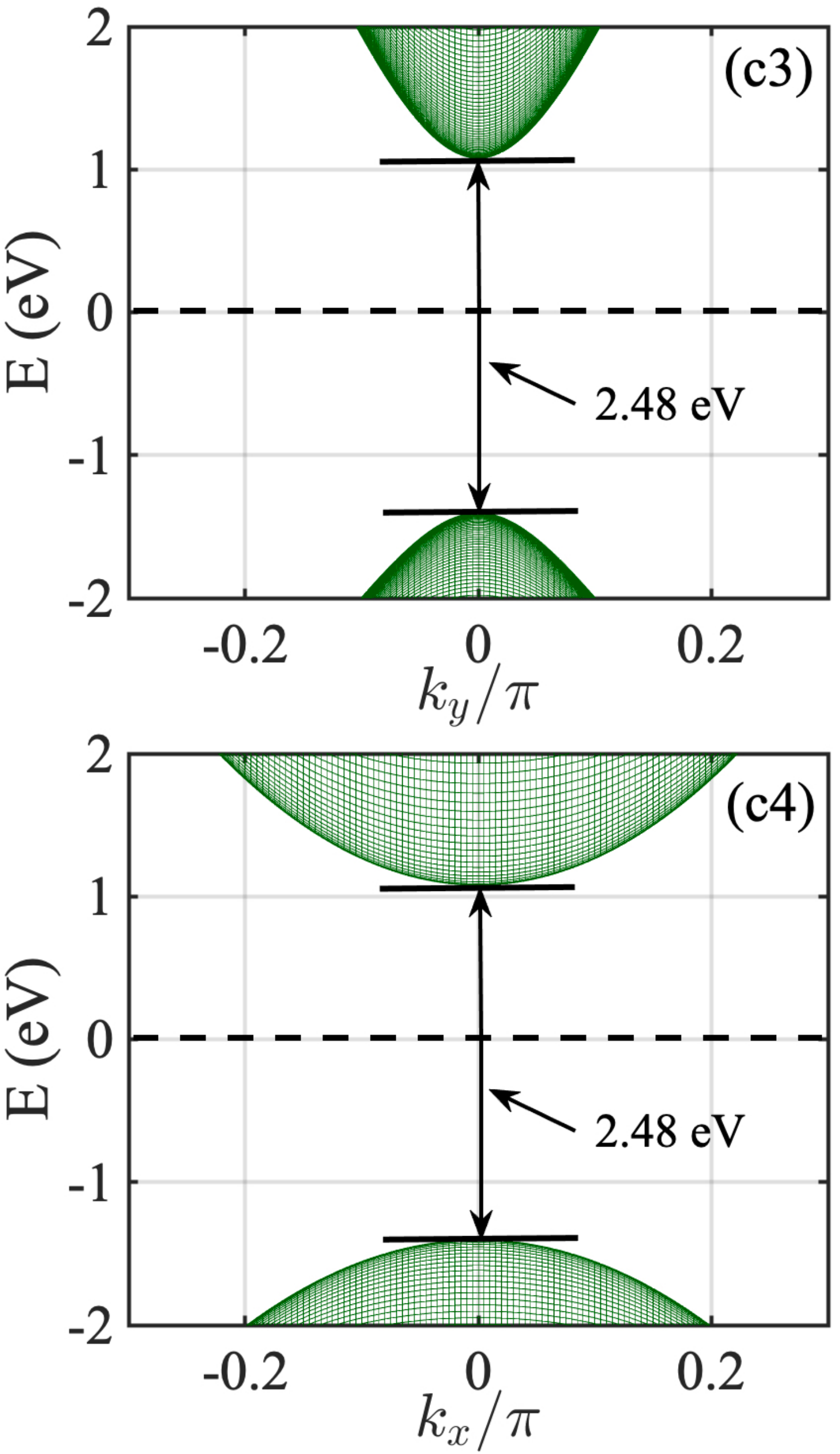}
\caption{\label{fig3}  The permittivity components and associated band structure 
when using the low-energy effective 
 Hamiltonian. The two top rows show 
 the $\epsilon_{xx}(\omega)$ and $\epsilon_{yy}(\omega)$ components, whereas the two bottom rows 
 depict the 
  band structure along the $k_x$ and $k_y$ directions. In columns (a), (b), and (c) biaxial in-plane strains
  are applied to the phosphorene system with
   strengths $-10\%$, $0\%$, and $+10\%$, respectively. }
\end{figure*}

\subsection{Optical transitions and band gap equivalence}\label{bandgap}

We begin with 
the
DFT-RPA approach to calculate the anisotropic dielectric response and band structure of phosphorene.
In Fig.~\ref{fig2},   
the permittivity components $\epsilon_{xx}(\omega)$ 
and $\epsilon_{yy}(\omega)$ are shown
as a function of frequency of the
 incident light.
The  associated band structure along the high-symmetry paths in \textbf{k}-space calculated by GPAW is also shown. 
 Biaxial in-plane strains of representative 
strengths $-10\%$, $0\%$, and $+10\%$ 
are applied to the
phosphorene in columns \ref{fig2}(a)-\ref{fig2}(c), respectively. 
The blue and red curves in the top and middle rows of Fig.~\ref{fig2} correspond 
to the real and imaginary parts
 of the permittivity components (as labeled). 
 Both  permittivity components $\epsilon_{xx}(\omega)$ and $\epsilon_{yy}(\omega)$
 at zero strain [Fig.~\ref{fig2}(b1) and Fig.~\ref{fig2}(b2)] are nonzero within the low-frequency regime, and approach
 zero at  $\omega\approx 4.7$~eV. 
 The imaginary parts of $\epsilon_{xx}(\omega)$ and $\epsilon_{yy}(\omega)$ 
 exhibit zero loss at frequencies below
 $\omega\approx 1.5$~eV and $\omega\approx 0.63$~eV, respectively. 
 As the onset of a nonzero imaginary permittivity generally points to  photon energies
 that generate electron interband transitions 
  in semiconductors and insulators, 
  one may conclude that the corresponding band structure of the results shown in Figs. \ref{fig2}(b1) and \ref{fig2}(b2) should possesses bidirectional band gap on the orders of $\omega\approx 1.5$~eV and $\omega\approx 0.63$~eV. 
  Note that
  the intraband transitions within the valence bands are not allowed due to the Pauli exclusion principle.
  Next, upon applying a compressive strain of $-10\%$,
   Figs.~\ref{fig2}(a1) and \ref{fig2}(a2) show that the real part of the
   permittivity now begins to diverge  when  $\omega\to 0$.
   There are also multiple zero crossings over the given frequency range, and
    peaks at $\omega\approx 0.59$~eV [Fig.~\ref{fig2}(a1)]
    and $\omega\approx 0.9$~eV [Fig.~\ref{fig2}(a2)]. 
    The different 
    threshold frequencies for 
    nonzero imaginary parts, i.e., $\omega\approx 0.45$~eV and $\omega\approx 0.55$~eV,
     suggest anisotropic interband transitions
     where the electronic transitions in the $x$ direction experience 
     a larger gap than those occurring in the $y$ direction. 
     Turning the strain type to tensile with the same magnitude,
     i.e., $+10\%$,
    Figs.~\ref{fig2}(c1) and \ref{fig2}(c2),
    show that  the permittivity components exhibit
     qualitatively  similar behavior to
      those of zero strain shown in Figs.~\ref{fig2}(b1) and \ref{fig2}(b2). 
      As is also seen, the frequency  thresholds where the imaginary parts vanish have increased to
       $\omega\approx 1.7$~eV and $\omega\approx 1.35$~eV, 
       compared to the cases with strains of
       $-10\%$ and $0\%$, suggesting that there is an increase in the energy gap for
        the interband transitions.   

To confirm the 
correlation between the band gap transitions and key regions of the frequency dispersion of the permittivity,
the band structure of 
 phosphorene along high symmetry paths in \textbf{k} space is plotted in Figs. \ref{fig2}(a3), \ref{fig2}(b3), and \ref{fig2}(c3). 
The phosphorene layer is subject to the same biaxial strain column-wise.
 The energies are scaled so that the Fermi level resides at $E=0$ (marked by the dashed line). 
 As seen in Fig.~\ref{fig2}(b3), 
 the unstrained system has a gap of $\sim 0.67$~eV at the $\Gamma$ point. It is known that the band
gap of phosphorene can be tuned by the number of layers, from $\sim$1.5~eV
in a monolayer to $\sim$0.59~eV in a five-stack layer.~\cite{bgclosing1,bgclosing2} Also, it was argued that the optical band gap of monolayer BP is around $\sim$1.5~eV, which is equivalent to a band gap of $\sim$2.3~eV
down-shifted by $\sim$0.8~eV through the  binding energy.~\cite{bgclosing1,bgclosing2} 
Note that to improve  band gap predictions, one can repeat the calculations with a hybrid functional,
 or make use of the GW approximation for the self-energy contribution.~\cite{GWA,PBE0,bgclosing1,bgclosing2}. 
Although it is known that the bare PBE functional underestimates the band gap of phosphorene 
($\sim1.52$~eV) \cite{Alidoust2020:PRB1},
  the information extracted earlier from the permittivity components calculated through DFT-PRA are not consistent with this band gap either. 
  By exerting $+10\%$ strain in Fig.~\ref{fig2}(c3), the band gap increases to $\sim 1.83$~eV, consistent with 
  the  behavior of the
   permittivity seen in Figs.~\ref{fig2}(c1) and \ref{fig2}(c2), 
   although the intricate 
    features that correlate with the interband transitions 
    at low energies are not consistent with the  band structure. 
  With   the application of $-10\%$ compressive strain, it is seen in Fig.~\ref{fig2}(a3) that
     the band gap closes and the conduction band at the $\Gamma$ point crosses the Fermi level. 
    Therefore, the associated permittivity should show metallic characteristics at low energies. 
    Indeed, the real part of permittivity in 
    Figs.~\ref{fig2}(a1) and \ref{fig2}(a2) acquires metallic properties with a 
    Drude-type response, centered around $\omega=0$, due to the 
    intraband transitions within the conduction band. 
    Nevertheless, the 
    imaginary part of permittivity in Figs.~\ref{fig2}(a1) and \ref{fig2}(a2) 
    does not overlap with the 
    Drude peak, suggesting an anisotropic band gap, which is incompatible with the band structure.   

We now discuss the finite-frequency optical conductivity and Drude response
within the framework of the low-energy model. To simplify our notation in what follows, we rewrite the low-energy Hamiltonian model (\ref{Hamil}) by introducing new parameters $a_{1,2},b_{1,2},c_{1,2}$:
\begin{align}
H(k_x,k_y)=& (a_1 + b_1 k_x^2 + b_2 k_y^2)\tau_0 \nonumber \\  
+ &(a_2 + c_1 k_x^2 + c_2 k_y^2)\tau_x - \chi_yk_y\tau_y.
    \label{Hamil2}
\end{align}
In this notation, the components of Green's function are given by 

\begin{subequations}\label{GF}
\begin{align}
G_{11,22}&(k_x,k_y, i\omega) = \nonumber \\&\frac{1}{2}\left( \frac{1}{i\omega -f_1+g_1} + \frac{1}{i\omega -f_1-g_1}\right),\\
G_{12,21}&(k_x,k_y, i\omega) = \nonumber \\&\frac{f_2\pm i\chi_yk_y}{2g_1}\left( \frac{1}{i\omega -f_1-g_1} - \frac{1}{i\omega -f_1+g_1}\right),
\end{align}
\end{subequations}
where the variables $f_1$, $f_2$, and $g_1$ are given by,
\begin{subequations}\label{GF}
\begin{align}
  f_1 &= a_1 + b_1 k_x^2 + b_2 k_y^2, \\
  f_2 &= a_2 + c_1 k_x^2 + c_2 k_y^2,\\
  g_1&=\sqrt{f_2^2+\chi_y^2k_y^2}.
\end{align}
\end{subequations}
Substituting the Green's function components (\ref{GF}) into Eq.~(\ref{optcond}), we obtain the real parts of the optical conductivity tensor, expressed in terms of Dirac-delta functions:
\begin{align}
    \sigma_{ab}(\omega) &= 
    \frac{e^2}{4 \pi \hbar \omega}
    \int  \int \int d\Omega \;dk_x \;dk_y {\cal F}(\Omega,\omega,\mu,T)\notag \\
     &\times \Big\{ h_{ab} \left[ 
    \delta(\Omega - f_1 + g_1) \delta(\omega + \Omega - f_1 - g_1) \right. \nonumber \\
    &\left.+ \delta(\Omega - f_1 - g_1) \delta(\omega + \Omega - f_1 + g_1)
    \right] \notag \\
    &+ g_{ab}^- 
    \delta(\Omega - f_1 + g_1) \delta(\omega + \Omega - f_1 + g_1)  \nonumber \\ 
    &\left.+ g_{ab}^+ \delta(\Omega - f_1 - g_1) \delta(\omega + \Omega - f_1 - g_1)
    \right]
    \Big\}.
    \label{eq:BP_sigma}
\end{align}
The temperature dependence of 
the optical conductivity in the continuum regime is given by 
${\cal F}(\Omega,\omega,\mu,T) = f(\Omega - \mu, T) - f(\Omega + \omega - \mu, T) $,
 in which $\mu$ stands for 
 the chemical potential and $f(X, T)$ is the Fermi-Dirac distribution at temperature $T$. 
 Also, we have introduced the following new variables to further simplify the final expressions
\small
\begin{subequations}
\begin{align}
    h_{xx} &= 4 k_x^2 \left( \frac{ c_1^2 \chi_y^2 k_y^2 }{f_2^2 + \chi_y^2 k_y^2} \right), 
\end{align}   
\begin{align} 
    g_{xx}^\pm &= 4 k_x^2 \left( b_1^2 + \frac{ c_1^2 f_2^2 }{ f_2^2 + \chi_y^2 k_y^2 } \pm \frac{2 b_1 c_1 f_2}{\sqrt{ f_2^2 + \chi_y^2 k_y^2 }} \right),  \end{align} 
\begin{align}
    h_{xy} &= 2c_1 k_x k_y\chi_y^2 \frac{2c_2k_y^2-f_2}{f_2^2 + \chi_y^2 k_y^2},
\end{align}   
\begin{align}    
    g_{xy}^\pm &= \frac{2 k_x k_y}{f_2^2 + \chi_y^2 k_y^2} \left( c_1f_2\pm b_1 \sqrt{f_2^2 + \chi_y^2 k_y^2}\right)\nonumber\\
     &\times \left( 2c_2f_2\pm 2b_2 \sqrt{f_2^2 + \chi_y^2 k_y^2} + \chi_y^2\right),  
\end{align}   
\begin{align}    
    h_{yy} &= \frac{ \left( f_2 - 2c_2 k_y^2 \right)^2 \chi_y^2 }{f_2^2 + \chi_y^2 k_y^2},  
\end{align}   
\begin{align}    
    g_{yy}^\pm &= k_y^2 \left(  4 b_2^2
    + \frac{\left( 2 c_2 f_2 + \chi_y^2 \right)^2}{ f_2^2 + \chi_y^2 k_y^2 }
    \pm \frac{ 4 b_2 \left(  2 c_2 f_2 + \chi_y^2 \right) }{ \sqrt{f_2^2 + \chi_y^2 k_y^2} } \right),
\end{align}   
\end{subequations}
\normalsize
Here $h_{yx}=h_{xy}$,
$g_{yx}^\pm=g_{xy}^\pm$, and the functions $f_1$ and $f_2$ are even functions of momenta $k_x$ and $k_y$. Therefore, $h_{ab}$ and $g_{ab}^\pm$ determine the symmetry of the optical
conductivity integrand [Eq.~(\ref{eq:BP_sigma})] with respect to momenta. 
As is seen, 
the integrands of $\sigma_{xy}(\omega)$ and $\sigma_{yx}(\omega)$ are odd functions 
of momenta due to the aforementioned symmetry properties of
 $h_{xy}, h_{yx}, g_{xy}^\pm$, and $g_{yx}^\pm$. 
 Hence, without performing any further calculations, we find that $\sigma_{xy}(\omega)=\sigma_{yx}(\omega)=0$ in this system. 
 On the other hand,
  $h_{xx}, g_{xx}^\pm, h_{yy}$, and $g_{yy}^\pm$ are even functions of momenta, and determine
   the 
 diagonal  optical conductivity tensor components 
 $\sigma_{xx}(\omega)$ and $\sigma_{yy}(\omega)$. 
The real part of the optical conductivity tensor, Eq.~(\ref{optcond}),
 is a complicated function of frequency and momenta that must be evaluated numerically. In what follows, 
 we first compute the optical conductivity as a function of $\omega$ and then
  obtain the components of permittivity through Eq.~(\ref{perm}).

In  Fig.~\ref{fig3}, a study  comparable to Fig.~\ref{fig2} is shown, except now we implement   the method based on
 the low-energy effective Hamiltonian,  Eq.~(\ref{Hamil}).
  The parameters for the
  Hamiltonian are obtained through fitting the model Hamiltonian to 
the band structure obtained  from
 first-principles around the $\Gamma$ point (summarized in Table \ref{table}). 
 The corrected band gap used in the low-energy Hamiltonian is on 
 the order of $1.52$~eV, although the magnitude of the
 band gap plays no role in our conclusions. 
 Comparing Figs.~\ref{fig2}(b1) and \ref{fig2}(b2) to Figs.~\ref{fig3}(b1) and \ref{fig3}(b2), 
 we see that the two approaches share similarities at low energies ($\omega \lesssim 3 \,{\rm eV}$). 
 For example, the generic 
  behaviors are similar, namely, $\epsilon_{xx}(\omega)$ has a flat, smooth variation with
   $\omega$ while $\epsilon_{yy}(\omega)$ has a clearly defined peak at low
    frequencies. Also, the strongly anisotropic nature of phosphorene is
    exhibited by the vastly different frequency-dependence of $\epsilon_{xx}(\omega)$
     and $\epsilon_{yy}(\omega)$. 
     Both DFT-RPA and the low-energy Hamiltonian model show some similar  trends, 
     i.e., the magnitudes follow $|\epsilon_{xx}(\omega)|$ $\ll$ $|\epsilon_{yy}(\omega)|$ within low energies. 
     There are however significant quantitative differences between the two approaches.
     The origins of these disagreements between the two are two-fold:
     First, the threshold value for nonzero 
   imaginary permittivity in both $\epsilon_{xx}(\omega)$ and $\epsilon_{yy}(\omega)$ obtained through the low-energy model are identical and equal to 
   $\omega=1.52$~eV, unlike the different values obtained using DFT-RPA. Hence
    the low-energy effective model suggests that the same band gap
    exists  in both directions. 
    Second, we find from  the low-energy model 
    that the
    real part of $\epsilon_{yy}(\omega)$ vanishes at $\omega=2$~eV despite the fairly large nonzero imaginary part of $\epsilon_{yy}(\omega)$ at the same frequency. This feature is absent in the DFT-RPA results and can play 
    a pivotal role in devising novel optoelectronics devices that are sensitive to loss.

Next, we incorporate strain, beginning  with a
 $+10\%$  tensile strain [Fig.~\ref{fig3}(c1) and Fig.~\ref{fig3}(c2)].
 It is observed that now
  the frequency cutoff for a  nonzero imaginary permittivity increases to $\omega=2.48$~eV, 
  suggesting an enlarged band gap. 
  Also, the permittivity is now nonzero over a larger interval of frequencies,
  indicating  a flattening of the
  conduction and valence bands following  application of this type of strain. 
  Reversing the strain direction,
  Figs.~\ref{fig3}(a1) and \ref{fig3}(a2) display the 
  permittivity components subject to $-10\%$ compressive in-plane strain. As seen, both the real and imaginary parts possess Drude absorption peaks, when $\omega\to 0$, indicating metallic behavior.
   Unlike the DFT-RPA results in Figs.~\ref{fig2}(a1) and \ref{fig2}(a2), 
   within the low-energy regime,
   the imaginary 
   component of the permittivity has a diverging Drude response for
    low frequencies,
     and a secondary peak 
      appears at $\omega=1.33$~eV and $\omega=1.30$~eV,
      for  $\epsilon_{xx}(\omega)$ and $\epsilon_{yy}(\omega)$, respectively. 

\begin{figure}[t!]
\includegraphics[width=0.45\textwidth]{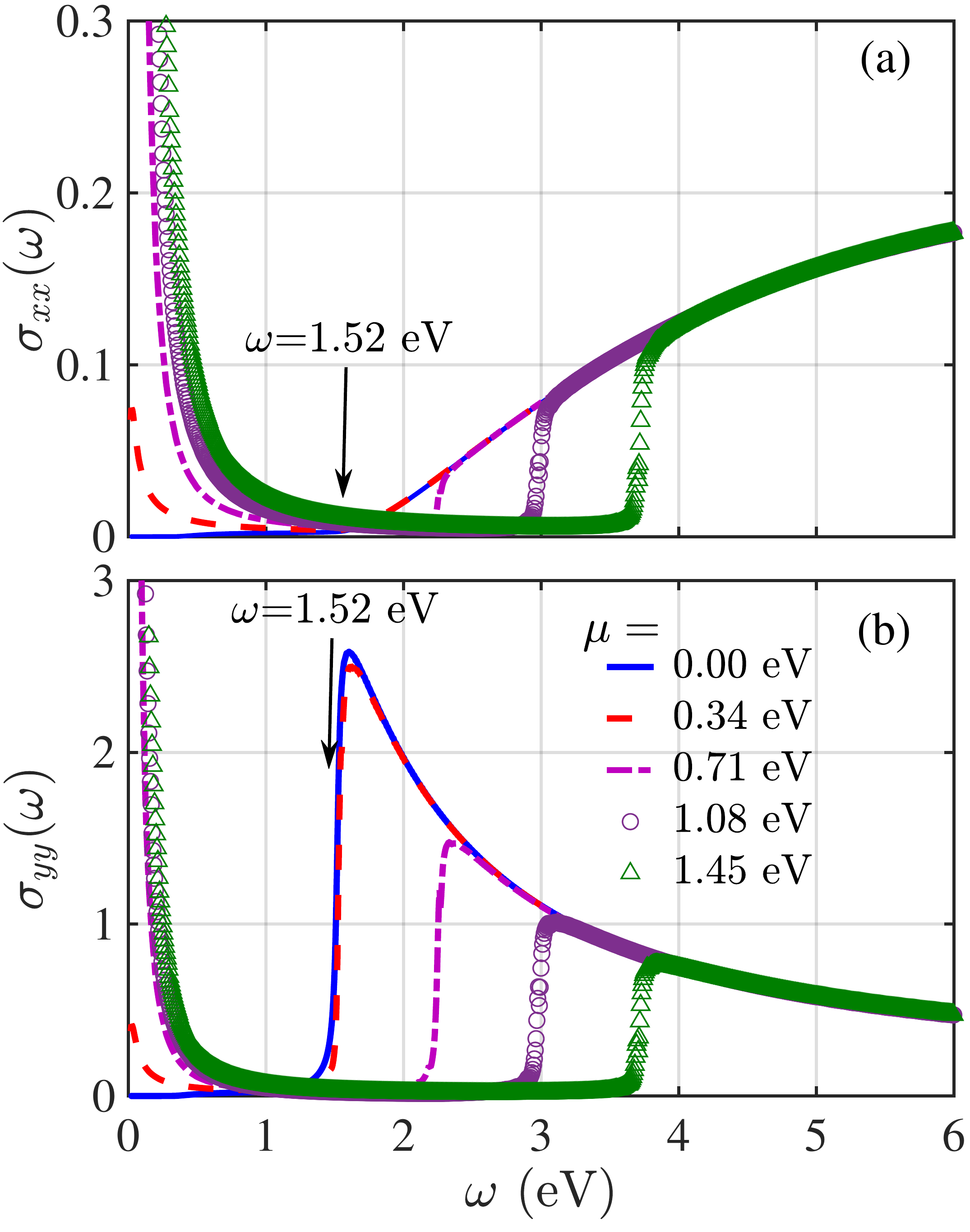}
\caption{\label{fig4} The normalized real part of 
the optical conductivity for a strain-free system obtained using the low-energy model. Panels 
(a) and (b) show the diagonal  components of the optical 
conductivity tensor as a function of frequency. 
The chemical potential takes the values, $\mu=0, 0.34, 0.71, 1.08, 1.45$~eV. }
\end{figure}

In order to fully understand these features, 
we have plotted the band structure associated with the
low-energy effective model along both the $k_x$ and $k_y$ directions. For the strain-free case,
the band structure in Figs.~\ref{fig3}(b3) and \ref{fig3}(b4) illustrates that the
 bottom of valence band and the top of conduction band are separated by a gap of $1.52$~eV for
  both directions.
  This
   clearly explains the identical  threshold values 
   for the 
nonzero imaginary permittivities 
shown in Figs. \ref{fig3}(b1) and \ref{fig3}(b2). The exertion of $+10\%$  tensile strain in Figs.~\ref{fig3}(c3) and \ref{fig3}(c4), 
 increases the band gap to $2.48$~eV for both the $k_x$ and $k_y$ directions, and 
 results in smaller band curvatures compared to unstrained phosphorene. 
 This accounts for  the
 nonzero imaginary permittivity
 for frequencies beyond the threshold  $\omega=2.48$~eV
 in Figs.~\ref{fig3}(c1) and \ref{fig3}(c2).
 For a 
 compressive strain of $-10\%$, 
  Figs.~\ref{fig3}(a3) and \ref{fig3}(a4) 
  show a closing of the gap, and the valence band now 
  crosses the Fermi level. 
  This crossing allows for
   intraband transitions, and thus
    the Drude peak for very low frequencies, $\omega\to 0$, emerges. 
    As seen, the band curvature now has further
    increased, resulting in a suppressed peak in the permittivity components [Figs.~\ref{fig3}(a1) and \ref{fig3}(a2)]. 
    Also, the two transitions at 
    the  
    energies of 
    $\omega=1.33$~eV and $\omega=1.30$~eV (marked in Figs.~\ref{fig3}(a3) and \ref{fig3}(a4))
  follow from  the 
  anisotropic band curvatures in the $k_x$ and $k_y$ directions. Note that one is unable to make an immediate conclusion for identifying the locations of these peaks by looking at the band structure because $\epsilon_{xx,yy}(\omega)$ are obtained by integrating over $k_x$ and $k_y$.    
  Unlike the zero strain and tensile strain cases,
   these differing transitions appear as peaks with differing locations in 
   both permittivity components presented
    in Figs.~\ref{fig3}(a1) and \ref{fig3}(a2). 
    
We have performed the DFT-RPA calculations and obtained
the permittivity components for a few materials and
semiconductors with moderate band gaps. Our results reveal
that the issues described here for the case of strained phosphorene also
appear for other material platforms.
This suggests that  the adverse effects inherited  from the RPA approach 
creates discrepancies that are generalizable to other systems.

\subsection{Optical conductivity and Drude weight}\label{Drude}

\begin{figure}[t!]
\includegraphics[width=0.4\textwidth]{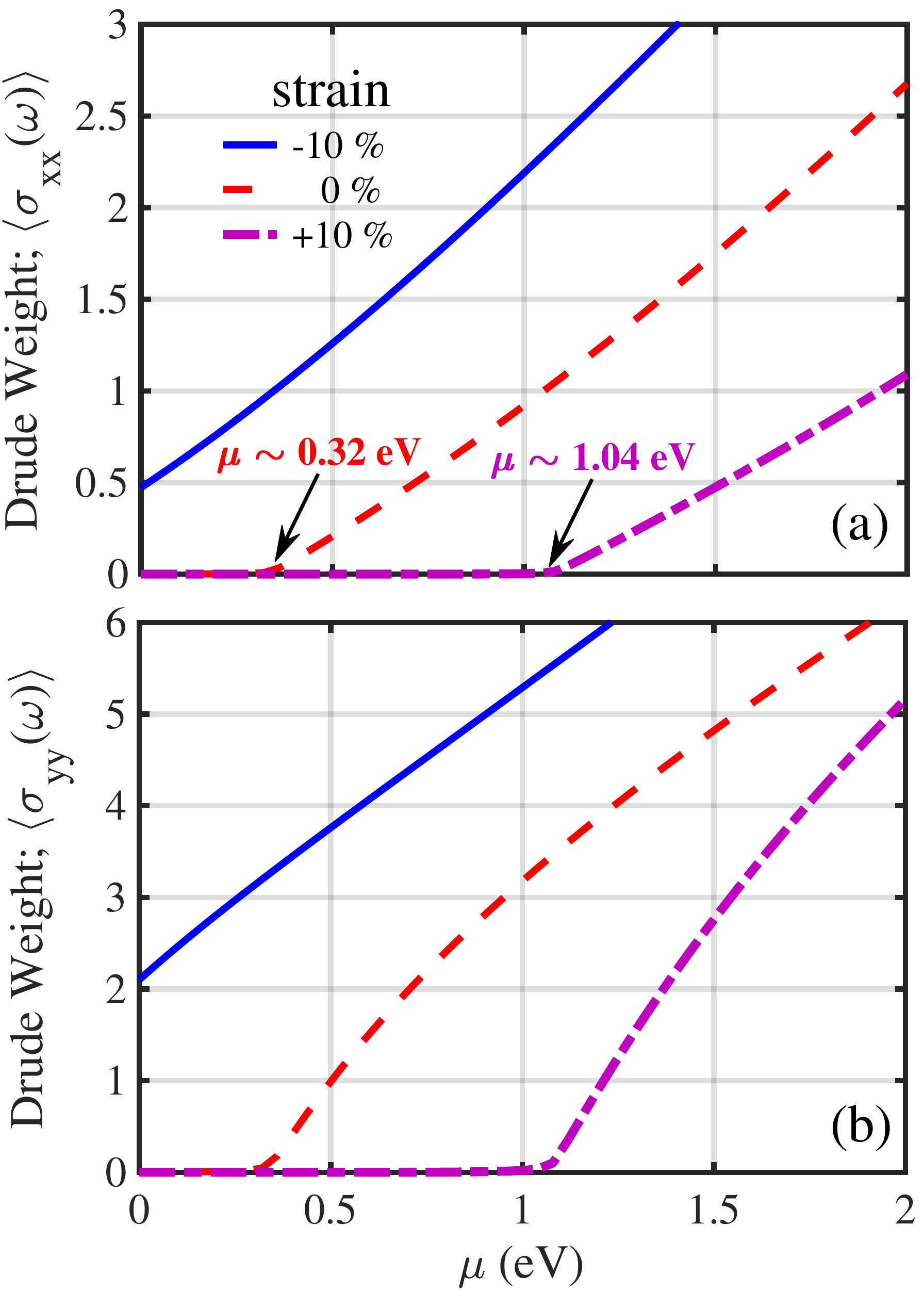}\\
\includegraphics[width=0.35\textwidth]{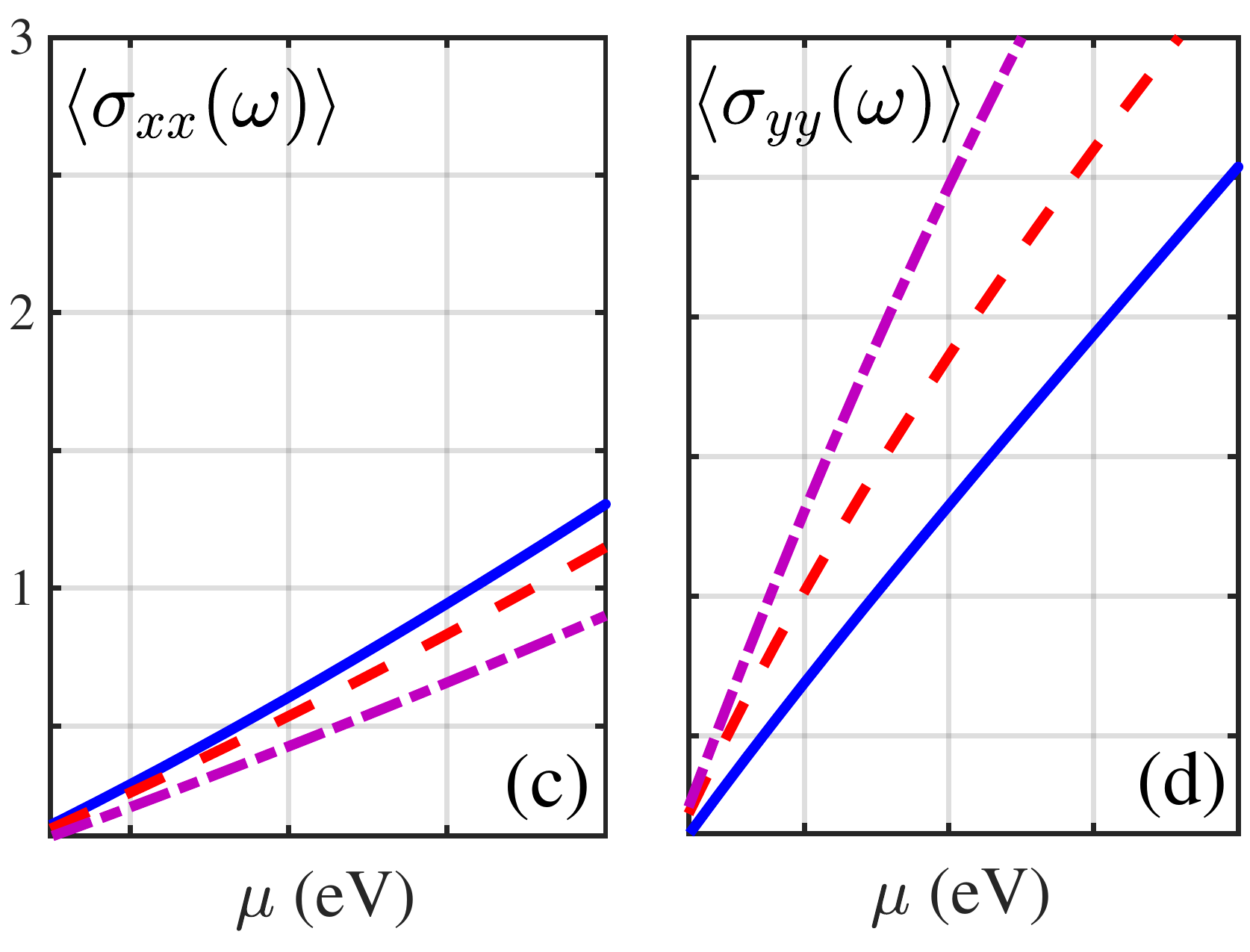}
\caption{
\label{fig5}  The Drude weight is shown as a function of chemical potential $\mu$. 
Three values of biaxial strain are considered: $-10\%, 0\%$, and  $+10\%$. 
The panels (a) and (b) show the Drude weight 
for
the conductivity components 
$\sigma_{xx}(\omega)$ and $\sigma_{yy}(\omega)$, respectively. 
For easier comparison,
 (c) and (d) show each curve from (a) and (b)  shifted to begin at the origin. 
 }
\end{figure}

For completeness, Fig.~\ref{fig4} presents the absorptive 
components of the
dynamical optical conductivity 
$\text{Re}\{ \sigma_{xx}(\omega)\}$ and $\text{Re}\{\sigma_{yy}(\omega)\}$ 
as a function of frequency in the strain-free system. 
We normalize each component by $\sigma_0=e^2/8\hbar$. To sample the different regions of the band structure, 
several representative values of the chemical potential are chosen. 
When the chemical potential is equal to zero or a value within the band gap, e.g., $\mu=0.34$~eV
(see Figs.~\ref{fig3}(b3) and \ref{fig3}(b4)),
 the  optical conductivity is zero at low frequencies and 
 then sharply rises at $\omega=1.52$~eV (the band gap magnitude),
  corresponding to the absorption onset. In other words, 
  the onset of nonzero optical absorption 
  is controlled by the band gap edges. 
  The associated transitions are schematically shown by arrows in Figs.~\ref{fig4}(a) and \ref{fig4}(b). By increasing the chemical potential, the Drude absorption peak persists as $\omega\to 0$, and the 
  onset of nonzero optical absorption shifts
   to higher values of $\omega$. 
  Note that when $\mu=0$ and $\mu=0.34$~eV,
  the optical conductivities  do not exhibit low-frequency divergences,
  as those energies reside inside the band gap. 
  Meanwhile, the small $\sigma_{xx,yy}(\omega)$ that is observed at low frequencies
 for
  $\mu=0.34$~eV is due to a
   small imaginary term $\eta=0.01$~eV, added to
    the frequencies for numerical stability, 
    and is physically 
    equivalent to nonelastic scattering. 
    When the chemical potential crosses the valence band at a representative value, e.g., $\mu=0.71$~eV, and larger values (see Figs. \ref{fig3}(b3) and \ref{fig3}(b4)), the Drude 
    response
     acquires
     more 
      pronounced values as $\omega\to 0$. 
      Similar to the components of 
      the permittivity tensor,
       the magnitudes of the components of
       the  optical conductivity tensor  obey
        $|\sigma_{xx}(\omega)| < |\sigma_{yy}(\omega)|$. 
      
        From observations of  the strain-free case,
        it is straightforward to understand how strain affects the dynamical optical conductivity.
        In the presence of e.g., $+10\%$ tensile strain,
        the  optical conductivity has the same structure as the strain-free case except now the band gap 
        increases to $2.48$~eV (see Figs.~\ref{fig3}(c3) and \ref{fig3}(c4)). Conversely, 
        a $-10\%$  compressive strain closes the band gap and, therefore, the low-frequency Drude response appears
        (even when $\mu=0$), with unequal dissipation threshold values, 
        i.e., $\omega=1.33$~eV and $\omega=1.30$~eV at the first peak of the optical conductivity 
        components $\sigma_{xx}(\omega)$ and $\sigma_{yy}(\omega)$, respectively. 
        This anisotropy originates again, from the differing curvatures of 
        the valence and conduction bands in different directions, 
        causing the 
        different interband transition gaps shown in Figs.~\ref{fig3}(a3) and \ref{fig3}(a4).             

The Drude weight of the anisotropic 
optical response can be obtained by integrating the Drude response part of 
the optical conductivity near $\omega\to 0$:  $
{\cal D}_{xx} = \lim_{\omega\to 0} \langle \sigma_{xx}(\omega)\rangle_\omega$, and 
${\cal D}_{yy}=\lim_{\omega\to 0}\langle\sigma_{yy}(\omega)\rangle_\omega$. 
Thus, ${\cal D}_{xx,yy}$ gives a weight to the zero-frequency divergence of the optical conductivity
and is closely associated with intraband transitions.
In Fig.~\ref{fig5}, we  
illustrate the Drude weight for  both components of
the optical conductivity $\sigma_{xx}(\omega)$ and $\sigma_{yy}(\omega)$ as a function of chemical potential $\mu$,
 for three biaxial strain values of
  $-10\%, 0\%$, and $+10\%$. 
 The calculations shown in Figs.~\ref{fig5}(a) and \ref{fig5}(b) reaffirm the 
  anisotropic Drude response in this system. 
  Nevertheless, the threshold chemical potential where the Drude weight
   becomes nonzero, is the same for both the $x$ and $y$ directions.
  Note that this threshold value for $\mu$,
  determines the distance between the Fermi level and the bottom of conduction band. When there is a
 compressive strain on phosphorene, 
 there is a Drude response, even at $\mu=0$. 
 The remaining strain cases 
 show that the distance between the Fermi level and the bottom of the conduction band is $0.32$~eV and $1.04$~eV
  in the presence of $0\%$ and $+10\%$ strain, 
  which are in excellent agreement with the band structure diagrams
   in Fig.~\ref{fig3}. 
   To illustrate the nonuniformity of the Drude response, 
   the Drude weight curves are shifted to
   the  origin in Figs.~\ref{fig5}(c) and \ref{fig5}(d). 
   As seen, the steepest response belongs to ${\cal D}_{yy}$ for the case of
   $+10\%$  strain, whereas 
   ${\cal D}_{xx}$ has a moderate response 
   for the same strain value. 
   This counterintuitive finding cannot be 
   deduced
    by simple examination of the conduction bands in Fig.~\ref{fig3}. 
    This is due to the fact that the band structure is
    nonuniform  in the $k_x \mbox{-} k_y$ plane 
    (see e.g., the isoenergy contour
     plots presented in Ref.~\onlinecite{Alidoust2019:PRB1}),
    and to obtain the Drude weight, the intraband transitions are integrated over the entire 
    isoenergy curves in the $k_x \mbox{-} k_y$ plane. 
    Nevertheless, to confirm these
     findings, we have performed checks by summing up the contributions from
     the  vertical intraband transitions through the following formula: \cite{kittel,Mahan} 
\begin{equation}
\lim_{\omega\to 0}\left\langle \sigma_{xx(yy)}(\omega) \right\rangle \propto \int \sum_{k_{x(y)}^i} \left| \frac{dE(k_{x(y)},k_{y(x)})}{dk_{x(y)}}\right|dk_{y(x)}.
\end{equation} 
As seen, this formula accounts for the vertical intraband transitions through the 
slope of the conduction band $E(k_x,k_y)$ at $E(k_x,k_y)=\mu$,
 where $k_{x(y)}^i$ are the roots of $E(k_x,k_y)-\mu=0$. The total Drude weight is proportional to the integration of all these states over $k_{y(x)}$. These calculations were found to have perfect agreement with 
 those presented in Fig.~\ref{fig5}. 

\begin{figure}[t!] 
\centering
\includegraphics[width=0.5\textwidth]{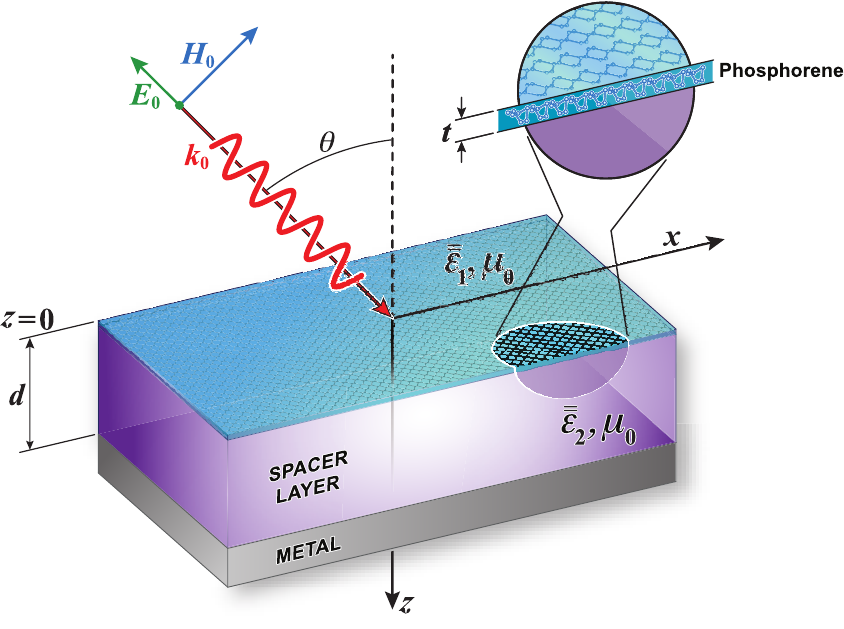}
\caption{Schematic of the configuration
  involving a  single-layer of  phosphorene
   with an effective thickness $t$ on
  top of a spacer layer of thickness $d$, and
metal reflecting substrate. The  phosphorene resides in the $x \mbox{-} y$ plane.
The phosphorene layer is exposed to an electromagnetic wave from the vacuum
region, where the incident electric field is polarized along $\hat{\bm{y}}$,
and the magnetic field is polarized in the $x \mbox{-}z$ plane.
The incident wavevector ${\bm k_0}$ makes an angle $\theta$ with the $z$ axis.
The crystallography principal directions are set as follows: 
$x\equiv a$, $y\equiv b$, and $z\equiv c$.
}
\label{fig6}
\end{figure}

\subsection{Implications for device design }\label{device}

We now
demonstrate 
that having an
accurate microscopic  model
for predicting the
optical response
of a material
is crucial for the successful 
design of even a simple  optics device, 
which in this case involves phosphorene.
As seen  in Fig.~\ref{fig6},
the basic design involves a
phosphorene layer 
deposited on top of an insulator layer with thickness $d$,
 and a perfect conductor, serving as a back plate. 
 The incident EM wave, propagating through vacuum, impinges on the device from the phosphorene side with an angle of $\theta$ 
 measured from the normal to the phosphorene plane.
In general, 
the permittivity tensor $\bar{\bar\varepsilon}$ 
and permeability tensor $\bar{\bar\mu}$ in the principal coordinates
take the following biaxial forms
 for a given (uniform) region ($n=0,1,2$):
 \begin{subequations}
\begin{align}
\bar{\bar\varepsilon}_n &= \varepsilon_{nx}\hat{\bm x}\hat{\bm x} + \varepsilon_{ny}\hat{\bm y}\hat{\bm y} + \varepsilon_{nz}\hat{\bm z}\hat{\bm z}, \\
\bar{\bar\mu}_n &= \mu_{nx}\hat{\bm x}\hat{\bm x} + \mu_{ny}\hat{\bm y}\hat{\bm y} + \mu_{nz}\hat{\bm z}\hat{\bm z},
\end{align}
\end{subequations}
where $n$ denotes the vacuum region ($n=0$), the phosphorene layer ($n=1$),
or the spacer layer ($n=2$). 

\begin{figure*}[t!]
\includegraphics[width=0.49\textwidth]{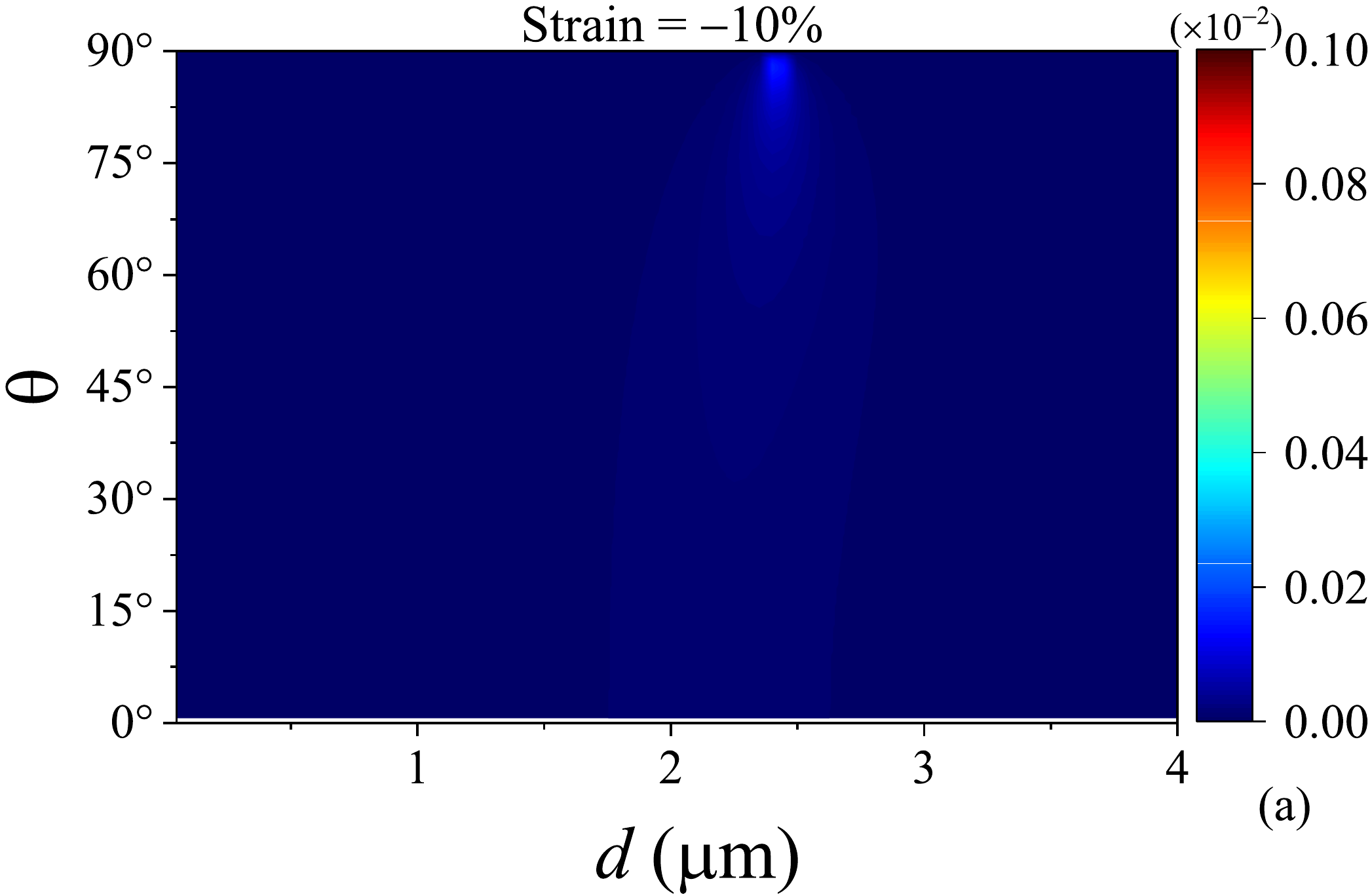}
\includegraphics[width=0.49\textwidth]{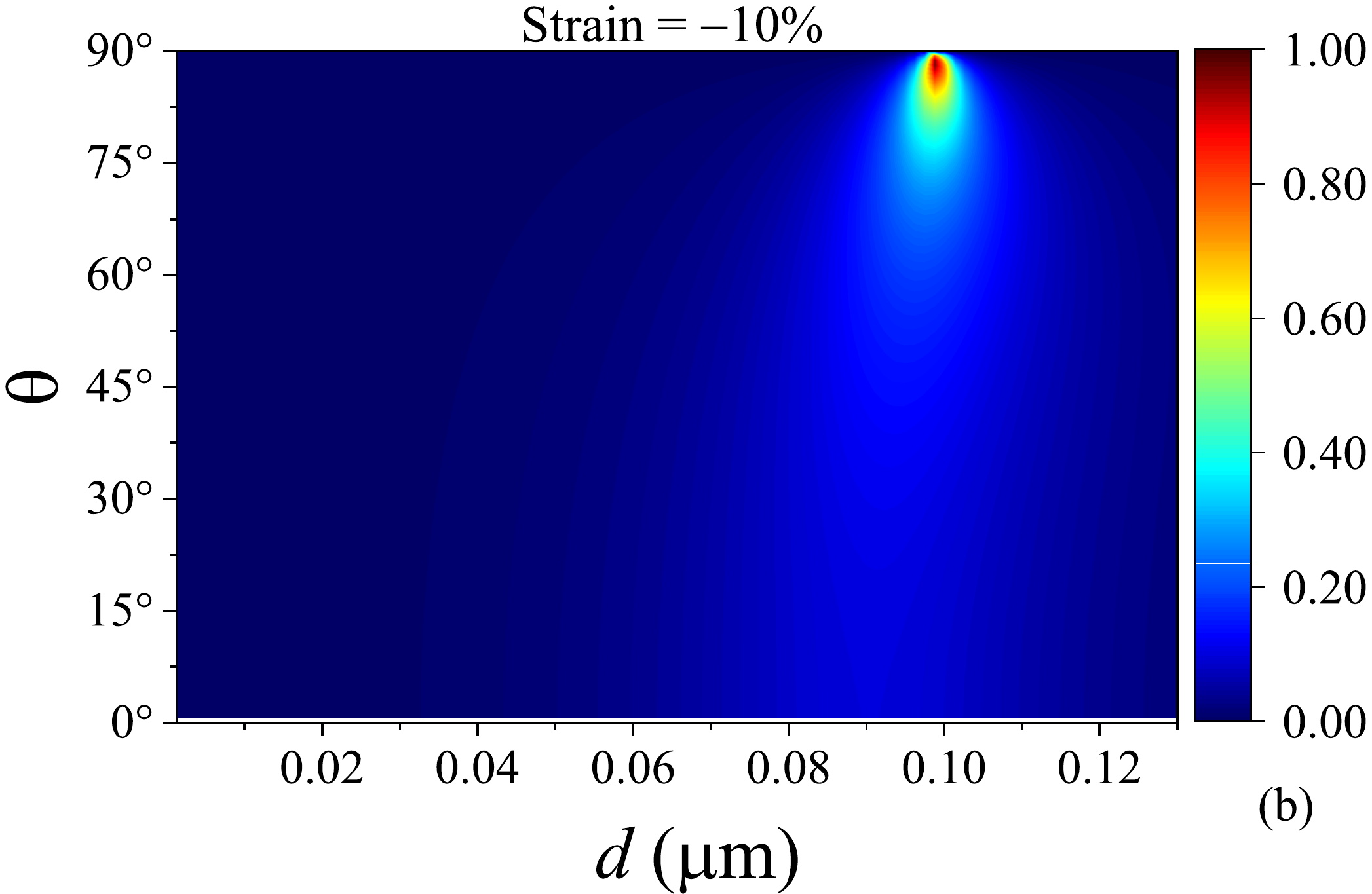}
\includegraphics[width=0.49\textwidth]{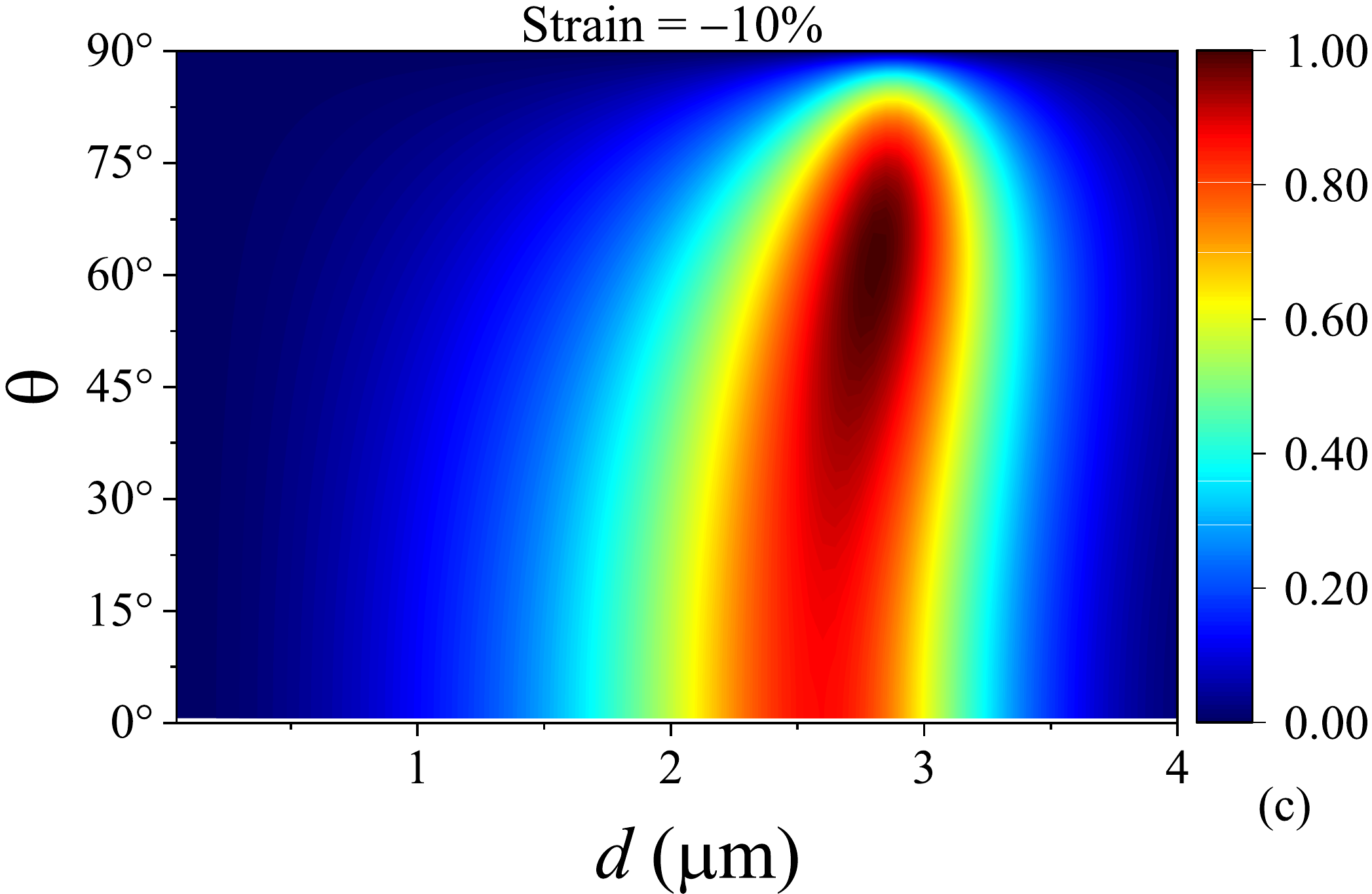}
\includegraphics[width=0.49\textwidth]{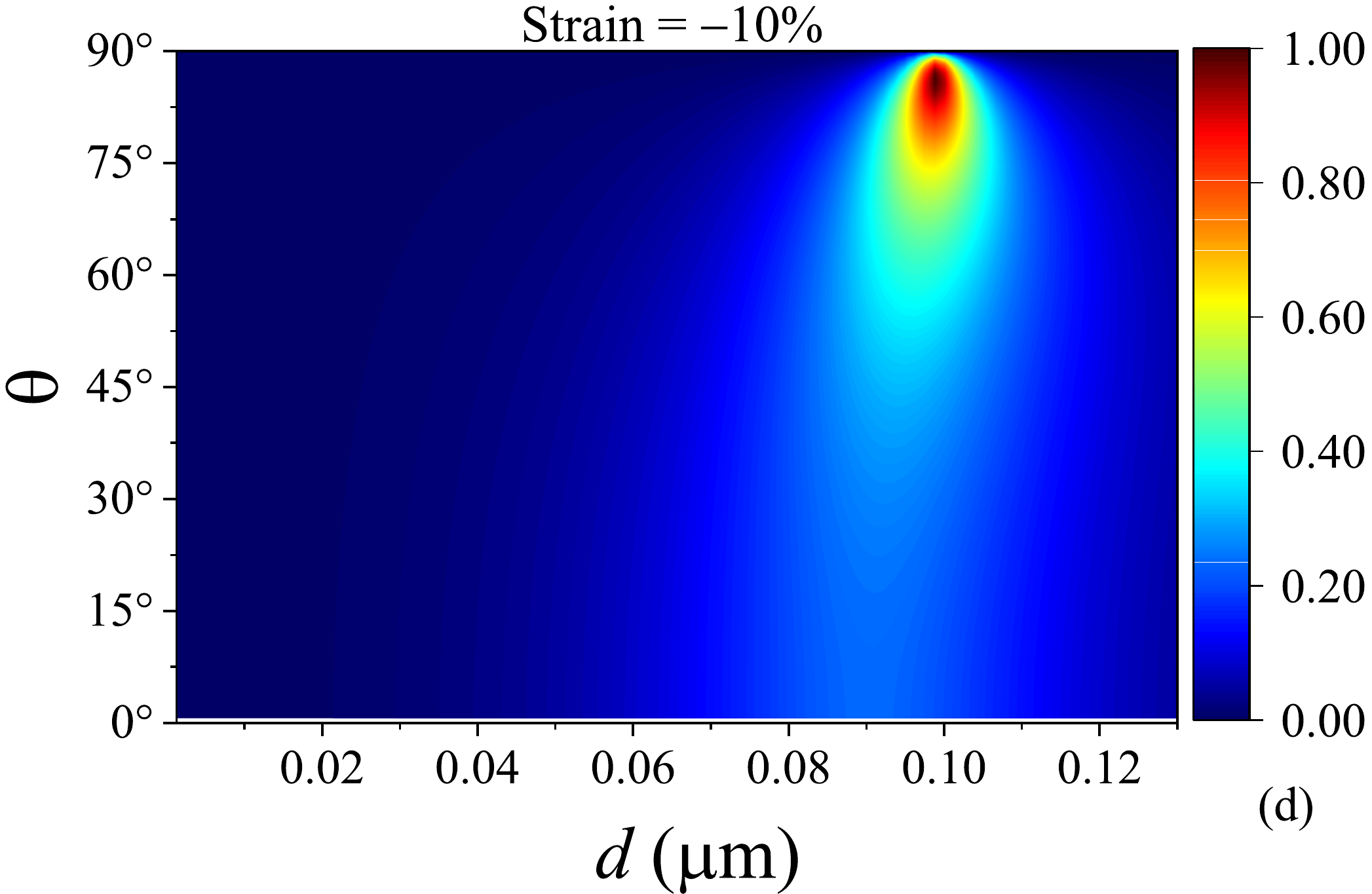}
\caption{\label{fig7} The absorptance as a function of the angle of incident light $\theta$ and the thickness of the insulator layer $d$.
 The DFT data and low energy model are used in the top and bottom rows, respectively.
  The frequency of the incident light is set to $\omega=0.06$~eV (left column) and $
  1.4$~eV (right column). 
  The biaxial strain is fixed at $-10\%$, and for proper comparisons, the chemical potential is set to $\mu=0$.   }
\end{figure*}

\begin{figure*}[t!]
\includegraphics[width=0.46\textwidth]{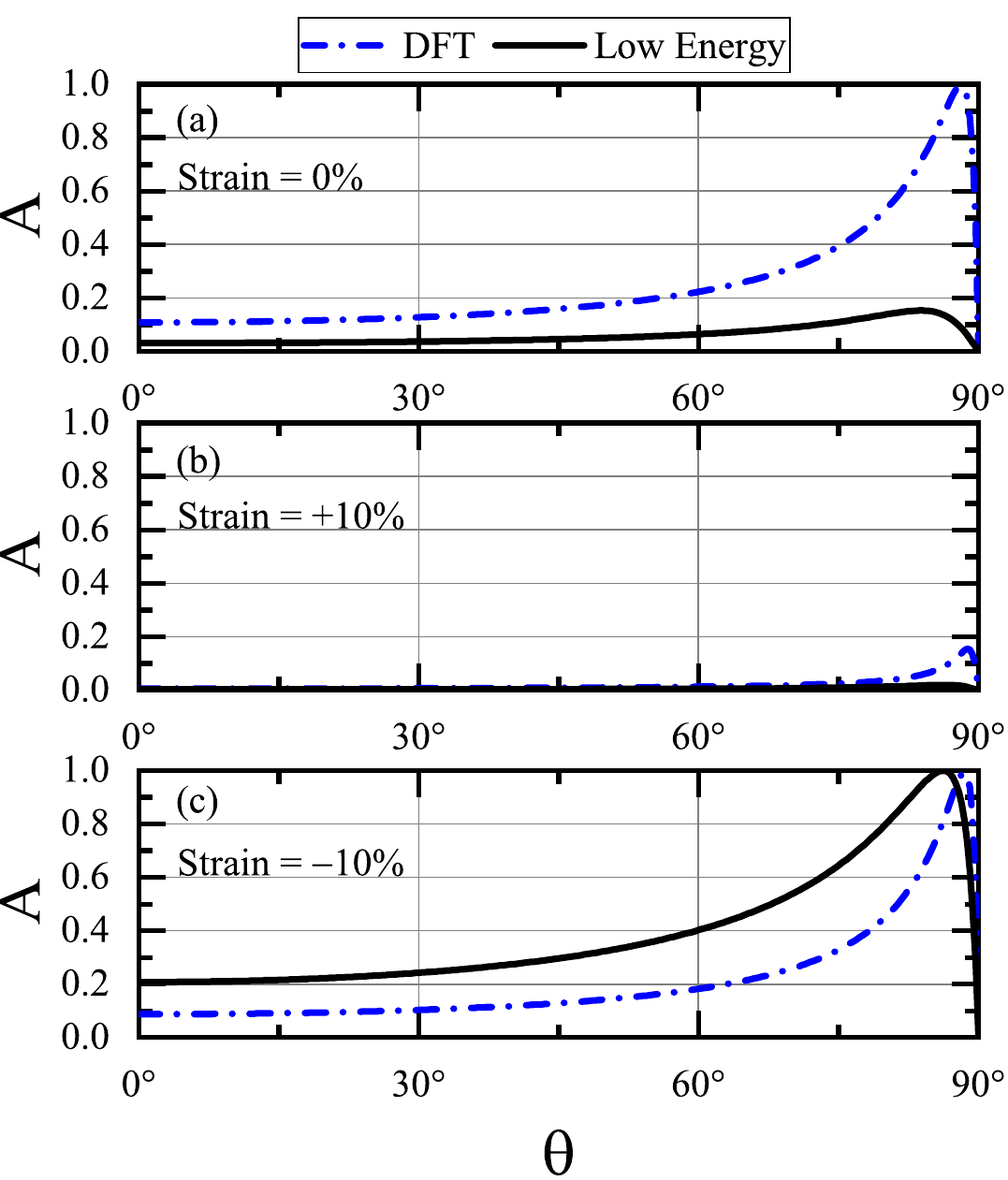}
\includegraphics[width=0.46\textwidth]{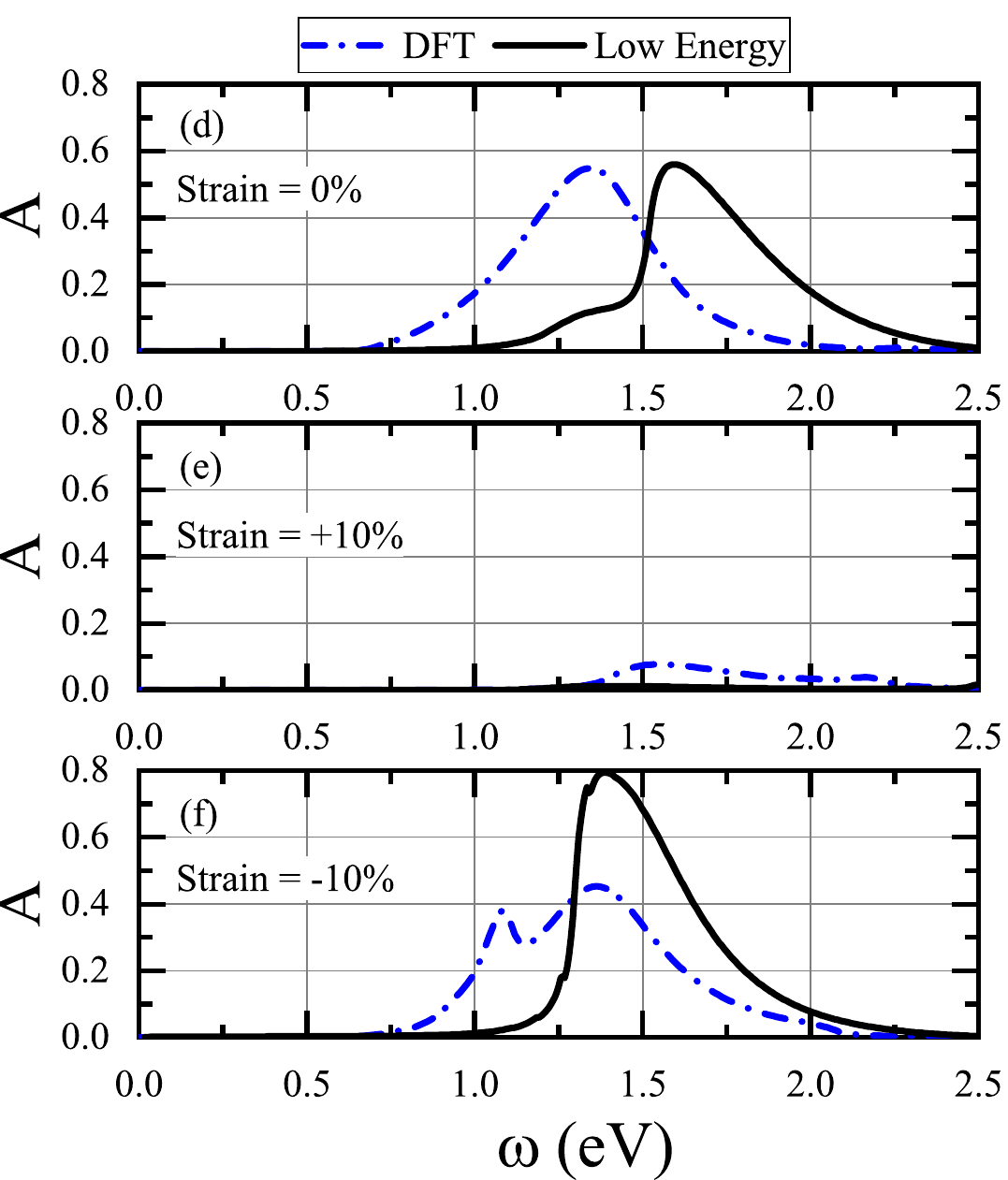}
\caption{\label{fig8} The absorptance $A$ calculated by employing the results of DFT and low-energy models. Three biaxial strain values are considered (as labeled): 
$-10\%$, $0\%$, $+10\%$. The left column shows $A$ as a function of the incident angle $\theta$,
at fixed frequency $\omega=1.4$~eV, and for an  insulator thickness of  $d=100$~nm.
The  right column has $A$ as a function of frequency of the incident EM wave at an incident angle of 
$\theta=80^\circ$, and $d=310$~nm.
 In all cases, the chemical potential is zero; $\mu=0$.  }
\end{figure*}

We now investigate the  absorption
of EM waves from 
the layered configuration shown in 
 Fig.~\ref{fig6}. The metallic back plate is taken to 
have perfect conductivity (PEC) for simplicity.
 The electric field of the  incident wave is polarized along $\hat{\bm{y}}$,
  and  is 
incident from the vacuum region 
  with wavevector ${\bm k}_0$  in the $x \mbox{-} z$ plane:
${\bm k}_0=\hat{\bm x}k_{0x} + \hat{\bm z} k_{0z}$,
where $k_{0x} = k_0 \sin\theta$, and $k_{0z} = k_0 \cos\theta$.
Since $\bar{\bar \varepsilon}_n$ has 
 no off-diagonal components, the transverse electric (TE) and transverse magnetic (TM) modes are decoupled.
The absorption can be calculated from Maxwell's equations.  
Assuming a harmonic time dependence $\exp(-i\omega t)$ for the EM field, 
we have,
\begin{eqnarray}
\label{d1}
\begin{split}
{\bm \nabla} \times {\bm E}_n &= +i \omega \mu_0 \bar{\bar\mu}_n {\cdot} {\bm H}_n ,\\
{\bm  \nabla} \times {\bm H}_n &= -i\omega \varepsilon_0 \bar{\bar \varepsilon}_n { \cdot} {\bm E}_n.
\end{split}
\end{eqnarray}
Combining Eqs.~(\ref{d1}), we get,
\begin{eqnarray}
\label{Maxw}
\begin{split}
{\bm \nabla}\times\bigl( \bar{\bar\mu}_n^{-1} \cdot {\bm \nabla}\times{\bm E}\bigr)  &=
\, k_0^2\bigl(\bar{\bar\varepsilon}_n \cdot{\bm E}\bigr)  \,,\\
{\bm \nabla} \times\bigl( \bar{\bar\varepsilon}_n^{-1} \cdot {\bm \nabla}\times{\bm H}\bigr)  &=
\, k_0^2\bigl(\bar{\bar\mu}_n \cdot{\bm H}\bigr).
\end{split}
\end{eqnarray}
We consider TE modes, corresponding to non-zero field components $E_{ny}$, $H_{nx}$, and $H_{nz}$.  
The electric field $E_{ny}$ satisfies
the following wave equation:
\begin{equation} \label{wave}
\frac{1}{\mu_{nz}} \frac{\partial^2 E_{ny}}{\partial x^2} + \frac{1}{\mu_{nx}} \frac{\partial^2 E_{ny}}{\partial z^2} + k_0^2\varepsilon_{ny} E_{ny} = 0  \,,
\end{equation}
which admits separable solutions of the form $\psi(z)\exp(i k_{0x} x)$.  
In what follows, we consider nonmagnetic media, so that $\mu_{nx}=\mu_{nz}=1$.
The parallel wave-vector $k_{0x}$ is determined by the incident wave, and is conserved across the interface. 
The form of $\psi(z)$ then simply  involves linear combinations of the exponential $\exp(i k_{nz} z)$ for
a given region. 
Thus,
 the electric  field 
in the vacuum region, ${\bm E}_0$, 
is written in terms of
 incident and reflected waves:
${{\bm E}_{0} \!=\!  (e^{i k_{0z} z} {+}  E_{0ry} e^{-i k_{0z} z})e^{i k_{0x} x} \hat{\bm y}}$.
From  the electric field, we can use Eq.~(\ref{d1}) 
to easily deduce
the magnetic  field components.
Due to the presence of the perfect metal plate,
in the spacer region, 
 the general form of the electric  field  is
 written in terms of standing waves:
${E_{2y}\!=\! E_2 \sin [k_2 (z-d)] e^{i k_{0x} x}}$, where
from Eq.~(\ref{wave}),
the wave number $k_2$ is given by,
 ${k_{2} \! =\! \pm \sqrt{\varepsilon_{2}k_0^2-k^2_{0x}}}$.
Note that 
 the boundary condition that $E_{2y}$ vanishes at the ground plane
 (${z\!=\!d}$) is
 accounted for (see Fig.~\ref{fig6}).
To construct the $\bm H$ fields
we  use Eqs.~(\ref{d1})
to arrive at,
${{\bm H}_n \! =\! (-\partial_z E_{ny},\partial_x E_{ny})/(i \eta_0 k_0)}$,
for $n=0,2$, and
where ${\eta}_0=\sqrt{\mu_0/\varepsilon_0}$ is the impedance of free space. 
The
presence of phosphorene enters in the boundary condition for the 
tangential component of the magnetic field by writing,
\begin{align} \label{currentbc}
\hat{\bf n} \times ({\bm H}_0-{\bm H}_2)=  {\bm J},
\end{align}
where $\hat{\bf n}$ is the normal to the vacuum/phosphorene interface, and ${\bm J}$ is the
current density.
Thus, we have $H_{2x}(z{=}0) {-} H_{0x}(z{=}0) {=}  J_y$,
where Ohm's law connects the surface current density $J_y$ in the phosphorene layer to the electric field in the usual way: 
${{\bm J} \!=\! \bar {\bar \sigma}{\bm E}}$. 
The dielectric tensor  in turn is defined through the 
surface conductivity tensor $\bar{\bar\sigma}$ via,
\begin{align} \label{surface}
\varepsilon_{ab}(\omega) \!=\! \delta_{ab} {+} \frac{i \sigma_{ab}(\omega)}{t \varepsilon_0 \omega},
\end{align}
where $t$ is the effective thickness of the phosphorene layer, which we take to be
$\approx {1\rm \, nm}$ (see Fig.~\ref{fig6}).
One can also consider the phosphorene layer as a finite sized slab,
like the spacer layer,
and solve for the fields within the layer.
This approach leads to equivalent results, but 
treating the phosphorene layer as a current sheet with  infinitesimal  
thickness
leads to simpler expressions.
Upon matching the tangential electric fields at the vacuum/spacer interface,
and using Eq.~(\ref{currentbc}),
it is straightforward to determine the unknown coefficients $E_{0ry}$
and $E_2$. 
First, the reflection coefficient $E_{0ry}$
is found to be,
\begin{align} \label{arr} 
E_{0ry}=-1 + \frac{2 \cos\theta }
{
i \kappa_2 \cot(k_2 d) + \cos\theta + \eta_0 \sigma_{yy} 
},
\end{align}
where we define $\kappa_2 = k_2/k_0$.
The coefficient $E_2$ 
for the electric field in the spacer region
is similarly found to be
\begin{align}
E_2=-\frac{2 \cos\theta \csc(k_2 d)}
{
 \cos\theta + \eta_0 \sigma_{yy} + i \kappa_2 \cot(k_2 d)
 },
\end{align}
where  from Eq.~(\ref{surface}), the dimensionless quantity 
$\eta_0 \sigma_{yy}$ can be expressed in terms of the permittivity:
$\eta_0 \sigma_{yy} =i (1 - \varepsilon_{yy}) k_0 t$.

The fraction of energy that is absorbed by the system is determined by the absorptance ($A$):  $A=1-R$,  
where $R$ is the reflectance index. Note that due to the metallic substrate, there is no transmission of EM fields into the region $z>d$.
In determining the absorptance of the phosphorene system, we
consider the time-averaged Poynting vector in the direction 
perpendicular to the interfaces (the $z$ direction), 
$S_{0z} {=} \Re{\{ {-} E_{0y} H_{0x}^*\}}/2$.
Upon inserting the electric and magnetic fields
for the vacuum region, we find,
${A \!=\! 1 \!-\! \left | E_{0ry} \right |^2}$, where 
${A \!=\! S_{z0}/S_0}$, and $S_0 \! \equiv \!  k_{0z}/(2 \varepsilon_0 \omega)$
 is the time-averaged Poynting vector for a
plane wave traveling 
in the $z$ direction.

In the following, we consider representative 
material and geometric parameters, and
demonstrate how 
the differing predictions from the
low-energy model and DFT-RPA can considerably  influence
the  absorption of EM energy in a phosphorene-based system.
We illustrate in
Fig.~\ref{fig7} the absorptance as a function of incident angle
 $\theta$, and 
 thickness of the insulator layer, $d$. The absorptance is determined after the
  permittivity tensor  is calculated from 
 the DFT-RPA and low-energy models.
 Results are shown in Figs.~\ref{fig7}(a) and \ref{fig7}(b) for the DFT-RPA approach,
  and in Figs.~\ref{fig7}(c) and \ref{fig7}(d), for the low-energy model.
  The frequency of the incident EM
  wave is set to $\omega=0.06$~eV in Figs.~\ref{fig7}(a) and \ref{fig7}(c), and 
  $\omega=1.4$~eV in Figs.~\ref{fig7}(b) and \ref{fig7}(d). 
  In all cases, 
  the chemical potential is set to zero and a compressive strain of $-10\%$ is considered.    
  Considering first the DFT-RPA results in \ref{fig7}(a) and \ref{fig7}(b), it is evident that over
  most of the parameter space,
  the incident beam  reflects completely off the structure ($A\approx 0$).
  There is only a small region of the diagram in \ref{fig7}(b) where there is near perfect absorption
  for near-grazing incidences ($\theta\approx 90^\circ$). The low-energy model in
Fig.~\ref{fig7}(c) however predicts an extremely strong 
 absorption region within $\rm 2\mu m\lesssim d \lesssim 3 \mu m$. 
 In contrast to the DFT approach, the low-energy model results in nearly perfect absorption within $\rm 45^\circ \lesssim \theta \lesssim 75^\circ$. Increasing the
 frequency to $\omega=1.4$~eV in Fig.~\ref{fig7}(d), the results of DFT and the
 low-energy model become more similar, although the low-energy model still predicts stronger absorption 
 over a broader range of  spacer layer thicknesses and angles $\theta$.

To further contrast the DFT-RPA method and low-energy model, 
Fig.~\ref{fig8} displays  the absorptance 
as a function of $\theta$ (left column)
and $\omega$ (right column), for three values of the biaxial
strain: $-10\%$, $0\%$, and  $+10\%$. Results in  the left column have a set frequency   $\omega=1.4$~eV, and an insulator thickness of $d=100$~nm,
whereas the right column has  $d=310$~nm and $\theta=80^\circ$.
Figure \ref{fig8}(c) corresponds to a slice of Figs.~\ref{fig7}(b) and \ref{fig7}(d), and more clearly shows 
how the DFT results cause a shifting of the near-perfect absorption peak towards $\theta=90^\circ$.
The discrepancies between the two models is seen to 
dramatically increase for the other strain values shown in \ref{fig8}(a) and \ref{fig8}(b).
Examining  the frequency response in  Figs.~\ref{fig8}(d)-\ref{fig8}(f), it is evident that 
the two models again lead to different absorption characteristics.
For zero strain [Fig.~\ref{fig8}(d)]  the absorptance profiles are similar but shifted in frequency.
When a tensile strain of $+10\%$ is applied to phosphorene [Fig.~\ref{fig8}(e)],
the DFT approach shows a small amount of absorption, but overall both models 
predict that the
 incident EM wave is mostly reflected over the given frequency window.
 When a compressive strain of $-10\%$ is applied [Fig.~\ref{fig8}(f)],
 there is again a shift similar to \ref{fig8}(d),
 but now the magnitudes of the peaks are much different, with 
 the low-energy model exhibiting near perfect absorption at $\omega=1.4$~eV.
 These discrepancies originate mainly
 from the different predictions for the permittivities, where e.g.,
 the DFT and low-energy methods give different 
 dissipation thresholds and
 significant amplitude variations (see Figs.~\ref{fig2} and \ref{fig3}).
 
 Regardless of the discrepancies and deviations discussed above (when using the permittivity data produced by the DFT-RPA and low-energy model), one can clearly observe the strong switching characteristics of the device considered (Fig.~\ref{fig6}) in absorbing the incident EM wave. As seen in Figs.~\ref{fig7} and \ref{fig8}, the device shown in Fig.~\ref{fig6} can absorb nearly perfectly the incident EM wave within certain incident angles, strain, the thickness of the insulator layer, frequency, and chemical potential. For example, comparing the results obtained for different values of strain in Figs.~\ref{fig8}(a,b,c), we conclude that the application of low strains (less than $10\%$) into the plane of phosphorene can effectively control the absorptivity of this device, switching efficiently between vanishingly small absorption and nearly perfect absorption of an incident EM wave at certain incident angles.
Although the overall behavior of the permittivity components in Figs.~\ref{fig2} and \ref{fig3} at 0$\%$ and +10$\%$ seem to be the same, at the given frequency, i.e., $\omega=1.4$~eV, these components possess substantially different imaginary and real parts, that together with the strong EM wave interference in the device of Fig.~\ref{fig6}, cause considerable differences in the absorptivity seen in Figs.~\ref{fig7} and \ref{fig8}. Both the interference phenomenon and Joule-heating effects are known to be strongly dependent on the amount of loss in the material and are governed by the imaginary part of the relevant permittivity component. For example, in the left set of figures in Fig.~\ref{fig7} (where the frequency is the same, and fixed at $\omega=0.06$~eV), the imaginary part of $\varepsilon_{yy}$ is: 0.00906 (DFT) and 157.4 (low-energy model). These huge variations in the dissipation translate into the observed absorptivity differences.

\section{conclusions}\label{conclusions}
Due to the fundamental importance of light-matter interactions,
we have  investigated the permittivity
 of phosphorene, subject to in-plane strain, as a representative material platform, using two approaches:
  One approach employed
   density functional theory combined with the 
   random phase approximation (DFT-RPA), and the other method involved
   a low-energy effective Hamiltonian model and Green's function. 
   The permittivity components for this strongly anisotropic material
   are fully explained by its associated band structures, transitions, and optical conductivities
   within the low-energy formalism. 
   However, the results of DFT-RPA and the
   corresponding band structures calculated from 
   the Perdew-Burke-Ernzerhof functional, showed considerable  discrepancies. 
   The DFT calculations were
    repeated using
     two different packages, and similar results were obtained. 
     Although some broad, generic trends for the frequency dispersion
    of the permittivity components 
     were in agreement for both approaches, several  important differences  
     stood out, including  
     the onset of the imaginary part of the  Drude response 
     that revealed important fundamental physical characteristics of a material, such as
     the  band gap, and interband and intraband transitions. 
     
     To illustrate the fundamental importance of accurate predictions of the permittivity response in designing new optoelectronics devices, we have compared the perfect absorption characteristics of a simple device, employing the permittivity data from the low-energy model and DFT-RPA.   
     Our results suggest that the DFT-RPA method implemented 
     in many types of  DFT packages 
     needs 
     to be revisited, and 
     improvements made so that 
     the results are at least more consistent with the associated band structures. 
     Accurate predictions for the
     permittivity and optical conductivity are of pivotal importance in determining the physical properties of materials and designing novel optoelectronics devices.
     
Interestingly, on the technological side of making use of phosphorene in optoelectronics devices, we find that strained phosphorene can serve as the switching element for the absorption of EM wave in EM wave absorbers. This switching effect can be effectively controlled by the application of relatively low mechanical strains (less than $10\%$) into the plane of phosphorene and/or manipulation of the chemical potential of phosphorene through a gate voltage.

\acknowledgements
The DFT calculations were performed using the resources provided by UNINETT Sigma2 - the National Infrastructure for High Performance Computing and Data Storage in Norway, NOTUR/Sigma2 project number: NN9497K. Part of the calculations were performed using HPC resources from the DOD High Performance Computing Modernization Program (HPCMP). K.H. is supported in part by the NAWCWD In Laboratory Independent Research (ILIR) program and a grant of HPC resources from the DOD HPCMP.

\end{document}